\documentclass[preprintnumbers,eqsecnum,prd,twocolumn]{revtex4}  

\usepackage[latin1]{inputenc}

\usepackage{graphicx}
\usepackage{latexsym}
\usepackage{amsmath}
\usepackage{amssymb}
\usepackage{amstext}
\usepackage{mathrsfs}

\def\beq{\begin{equation}}
\def\eeq{\end{equation}}
\begin{document}

\title{Detweiler's redshift invariant for extended bodies orbiting a Schwarzschild black hole}

\author{Donato Bini$^{1,2}$, Andrea Geralico$^1$, and Jan Steinhoff$^3$}
  \affiliation{
$^1$ Istituto per le Applicazioni del Calcolo ``M. Picone,'' CNR, I-00185 Rome, Italy\\
$^2$ INFN, Sezione di Roma Tre, I-00146 Rome, Italy\\
$^3$ Max Planck Institute for Gravitational Physics (Albert Einstein Institute), Am M\"uhlenberg 1, Potsdam 14476, Germany
}

\date{\today}

\begin{abstract}
We compute the first-order self-force contribution to Detweiler's redshift invariant for extended bodies endowed with both dipolar and quadrupolar structure (with spin-induced quadrupole moment) moving along circular orbits on a Schwarzschild background.
Our analysis includes effects which are second order in spin, generalizing previous results for purely spinning particles.
The perturbing body is assumed to move on the equatorial plane, the associated spin vector being orthogonal to it.
The metric perturbations are obtained by using a standard gravitational self-force approach in a radiation gauge.
Our results are accurate through the 6.5 post-Newtonian order, and are shown to reproduce the corresponding post-Newtonian expression for the same quantity computed by using the available Hamiltonian from an effective field theory approach for the dynamics of spinning binaries.
\end{abstract}

\maketitle

\section{Introduction}

The detection of the first binary neutron star inspiral by the LIGO-Virgo interferometers \cite{TheLIGOScientific:2017qsa}, which was likely already accompanied by a second one \cite{Abbott:2020uma} and is expected to be followed by hundreds of similar events during the next observing runs, has provided us an unique opportunity for improving our knowledge about the internal structure of neutron stars and the equation of state of neutron star matter.
The analysis of the associated gravitational wave signal has allowed one to impose tight constraints on the component masses, spins and tidal polarizability parameters as well as to measure their radii and equation of state \cite{Abbott:2018wiz,Abbott:2018exr}.

Spin effects may significantly modify both the orbital motion and the rate of the inspiral, since each neutron star gets deformed due to its own rotation \cite{Poisson:1997ha}.
As a result, such a spin-induced quadrupole moment introduces additional variations in the emitted gravitational wave signal, which are expected to dominate with respect to tidal effects in the case of rapidly rotating neutron stars \cite{Harry:2018hke}.
Furthermore, the quadrupole moment of a rotating neutron star is different from that of a spinning black hole, depending on the equation of state \cite{Laarakkers:1997hb}, so that any deviation from the black hole value can be used to constrain the binary black hole nature of the compact binary system \cite{Krishnendu:2017shb,Krishnendu:2019tjp}.

Finite size effects on the motion of two bound compact objects are taken into account in the literature by a number of different methods and at different levels of approximation.
At the lowest level, the dynamics of an extended body in a given gravitational background field is commonly described according to the Mathisson-Papapetrou-Dixon (MPD) model~\cite{Mathisson:1937zz,Papapetrou:1951pa,Dixon:1970zza}. The orbit is no longer geodesic due to the coupling between spin and higher multipole moment tensors with the background curvature tensor and its derivatives, and the spin vector is no more Fermi-Walker transported along the orbit due to the same type of couplings.
Such a feature complicates the discussion of the motion, which already at this simplest level cannot be performed exactly, but only within some approximation scheme and under some simplifying assumption \cite{Bini:2008zzc,Bini:2008zzf,Steinhoff:2009tk,Steinhoff:2012rw,Bini:2013nw,Bini:2013uwa,Bini:2014xyr,Bini:2014epa,Bini:2015zya}.
A canonical Hamiltonian formulation of the dynamics of a spinning test particle in a curved spacetime has been developed, e.g., in Ref. \cite{Barausse:2009aa} (see also Refs. \cite{Kibble:1963,Khriplovich:1989kg,Witzany:2018ahb}), later generalized to extended bodies endowed with spin-induced quadrupole moment in Ref. \cite{Vines:2016unv}.

When the mass of the extended body cannot be considered as a test mass, backreaction effects cannot be neglected, and the situation worsens immediately.
Analytical methods are still available: post-Newtonian (PN) \cite{Blanchet:2013haa,Schafer:2018kuf,Pati:2000vt,Futamase:2007zz} and  post-Minkowskian (PM) \cite{Bel:1981be,Westpfahl:1985} approximations, for arbitrary values of the mass ratio, possibly implemented using effective field theory (EFT) techniques \cite{Goldberger:2007hy,Rothstein:2014sra,Cheung:2018wkq}; the gravitational self-force (GSF) formalism \cite{Detweiler:2008ft,Barack:2009ux,Bini:2013zaa}, valid in the extreme-mass-ratio limit. The formalism which encompasses all these approaches is nowadays the effective-one-body (EOB) model \cite{Buonanno:1998gg,Buonanno:2000ef}, which is currently used to build waveform models for LIGO and Virgo data analysis.
It represents the most versatile framework which allows one to convert information coming from both analytical approaches and numerical relativity (NR) simulations
of binary inspirals to provide even more accurate predictions for the analysis of gravitational wave signals.

Up to now the PN description of the conservative orbital features of a two-body system is at the 4PN level of accuracy \cite{Damour:2014jta,Damour:2015isa,Bernard:2016wrg,Bernard:2017ktp,Foffa:2019yfl,Blumlein:2020pog}.
The spin part of the conservative dynamics is complete to 4.5PN order (for rapidly rotating compact objects), which includes next-to-next-to-next-to-leading-order (NNNLO) effects at the spin-orbit level \cite{Antonelli:2020aeb} (see also Ref.~\cite{Levi:2020kvb}), next-to-next-to-leading-order (NNLO) effects at the spin-squared level \cite{Hartung:2011ea,Levi:2011eq,Hartung:2013dza,Levi:2014sba,Levi:2015ixa,Levi:2015msa,Levi:2016ofk}, next-to-leading-order (NLO) at cubic order in spin \cite{Levi:2019kgk} (see also Refs. \cite{Siemonsen:2019dsu}), and leading-order (LO) effects at quartic order in spin \cite{Levi:2014gsa} (see also Refs. \cite{Hergt:2007ha,Hergt:2008jn,Vaidya:2014kza}).
The PM description has provided recently a 3PM orbital Hamiltonian \cite{Bern:2019nnu,Bern:2019crd,Cheung:2020gyp,Blumlein:2020znm} and the leading PM order in the spin part for black holes \cite{Vines:2017hyw,Chung:2020rrz} (with the spin-orbit part being universal).
Several orbital invariants have been computed within the GSF approach through a very high PN accuracy, but only in the case of particles without internal structure (see, e.g., the review \cite{Barack:2018yvs} and references therein).
Recently, linear-in-spin corrections to the Detweiler's redshift invariant and to tidal invariants have been obtained in Refs. \cite{Bini:2018zde,Bini:2018svh} for spinning particles moving along circular orbits in a Schwarzschild spacetime.
Finally, an improved description of spin effects in the EOB Hamiltonian has been presented in Ref. \cite{Khalil:2020mmr}, valid for arbitrary spin orientations and magnitudes, and for either black holes or neutron stars (see Ref. \cite{Nagar:2018plt} for the case of aligned spins).

In this work we compute high-order PN corrections to Detweiler's redshift invariant due to an extended body with spin-induced quadrupolar structure moving along circular equatorial orbits on a Schwarzschild background.
These corrections are quadratic in the spin of the body.
The spin vector is taken orthogonal to the motion plane, and generates a quadrupole moment which is proportional to a \lq\lq polarizability'' parameter normalized in such a way that it equals unity in the case of a black hole, whereas for neutron stars it depends on the equation of state.
The associated energy-momentum tensor used as the source of the first-order perturbation equations is highly-singular at the body's position, containing Dirac-delta terms as well as both first and second derivatives of the Dirac-delta function. We follow the standard Teukolsky approach and the Chrzanowski-Cohen-Kegeles (CCK) procedure to reconstruct the metric perturbation in a radiation gauge \cite{Teukolsky:1973ha,Cohen:1974cm,Chrzanowski:1975wv,Kegeles:1979an,Sasaki:2003xr,Keidl:2010pm}
. We introduce a dimensionless spin parameter to control the spin and spin-squared part of the perturbation, in a consistent way, all along the various steps that the Teukolsky formalism requires: source term, PN-type and Mano-Suzuki-Takasugi (MST) type solutions \cite{Mano:1996vt,Mano:1996mf,Mano:1996gn} of the radial homogeneous equation, Green-function, etc.
There are also \lq\lq side problems," like that of determining the nonradiative part of the perturbation  associated with the \lq\lq low multipoles,"  or better---in the context of the Teukolsky approach---the gauge-dependent mass and angular momentum perturbations due to the extended body. We solve this problem here by following the same approach already used in our previous work valid to linear order in the particle's spin \cite{Bini:2018zde} within the Regge-Wheeler-Zerilli (RWZ) \cite{Regge:1957td,Zerilli:1971wd} framework.
Finally, the result for the redshift is compared to PN predictions, thereby providing the first independent check of the NNLO PN spin-squared potential \cite{Levi:2015ixa}.

We use geometrical units $G=1=c$.
Greek indices refer to spacetime coordinates and vary from 0 to 3, whereas Latin indices, ranging from 1 to 3, label space coordinates.

\section{MPD description of quadrupolar bodies}

The motion of an extended body endowed with structure up to the quadrupole in a given spacetime is described by the MPD equations \cite{Mathisson:1937zz,Papapetrou:1951pa,Dixon:1970zza}
\begin{eqnarray}
\label{eq_MPD}
\frac{DP^\mu}{d\tau} &=& -\frac12 R^\mu{}_{\nu\alpha\beta}U^\nu S^{\alpha\beta}-\frac16 J^{\alpha\beta\gamma\delta}\nabla^\mu R_{\alpha\beta\gamma\delta}\nonumber\\
& \equiv&  F^\mu_{\rm (spin)} + F^\mu_{\rm (quad)} 
\,,\nonumber\\
\frac{DS^{\mu\nu}}{d\tau} &=& 2P^{[\mu} U^{\nu]}+\frac43 J^{\alpha\beta\gamma[\mu}R^{\nu]}{}_{\gamma\alpha\beta}\nonumber\\
&\equiv&  D^{\mu \nu}_{\rm (spin)} + D^{\mu \nu}_{\rm (quad)}
\,,
\end{eqnarray}
where 

\begin{enumerate}

\item $U=\frac{dx^\alpha}{d\tau}\partial_\alpha$ is the (timelike, $U\cdot U=-1$) unit tangent vector to the \lq\lq center of mass world line'' (${\mathcal C}$, with parametric equations $x^\alpha=x^\alpha(\tau)$) used to make the multipole reduction, parametrized by the proper time $\tau$.

\item $P=m u$, with $u\cdot u=-1$ and $P\cdot P=-m^2$, is the (timelike) generalized 4-momentum of the body with mass $m$. Note that, in general, $U$ and $u$ are not aligned; $P$ (i.e., $u$) has support only along ${\mathcal C}$; $m$ does not coincide with the \lq\lq bare mass'' of the body, but depends on its structure.

\item $S^{\mu\nu}$ is a antisymmetric spin tensor $S^{\mu\nu}$ (with support only along ${\mathcal C}$, like $P$), which is assumed to satisfy the Tulczyjew-Dixon supplementary conditions~\cite{Dixon:1970zza,tulc59}
\beq
\label{TDconds}
S^{\mu\nu}u_\nu=0\,.
\eeq
As standard, the spin vector (orthogonal to $u$) associated with the spin tensor $S^{\alpha\beta}$ is given by
\begin{eqnarray}
S(u)^\alpha&=&\frac12 \eta(u)^{\alpha\beta\gamma}S_{\beta\gamma}\,,
\end{eqnarray}
where $\eta(u)_{\alpha\beta\gamma}=u^\mu \eta_{\mu\alpha\beta\gamma}$
is the spatial unit volume 3-form (with respect to $u$) built from the unit volume 4-form
$\eta_{\alpha\beta\gamma\delta}=\sqrt{-g}\, \epsilon_{\alpha\beta\gamma\delta}$,
with $\epsilon_{\alpha\beta\gamma\delta}$ ($\epsilon_{0123}=1$) being the
Levi-Civita alternating symbol and $g$ the determinant of the metric.

Its signed magnitude $s$ is such that
\begin{eqnarray}
s^2&=&S(u)\cdot S(u)=\frac12 S_{\mu\nu}S^{\mu\nu}=-\frac12 {\rm Tr}[ S^2]\,,
\end{eqnarray}
with $[S^2]^\alpha{}_\beta =S^{\alpha\mu}S_{\mu\beta}$, and is not constant in general along the trajectory of the extended body.
For a later use, it is convenient to introduce the symmetric-tracefree part (STF) of the square of the spin tensor (or, equivalently, of the spin vector) $S^2$, i.e.,
\beq
[S^2]^{\rm STF}{}^{\alpha\beta}=[S^2]^{\alpha\beta}-\frac13 P(u)^{\alpha\beta} {\rm Tr}[ S^2]\,,
\eeq
where $P(u)=g+u\otimes u$ projects orthogonally to $u$. One finds 
\beq
[S^2]^{\rm STF}=[S(u)\otimes S(u)]^{\rm TF}\,.
\eeq

\item $J^{\alpha\beta\gamma\delta}$ is the quadrupole tensor, with support only along ${\mathcal C}$, like $P$ (and $u$) and $S^{\mu\nu}$ (and $S(u)$).
It shares the same symmetries of the Riemann tensor and is completely specified by two symmetric and trace-free spatial tensors, i.e., the mass quadrupole (electric) and the current quadrupole (magnetic) tensors \cite{Dixon:1970zz,ehlers77,Bini:2013nw,Bini:2013uwa,Bini:2014xyr}.

\end{enumerate}

We will consider here the case of a spin-induced quadrupole tensor of the electric-type only, i.e.,
\beq
J^{\alpha\beta\gamma\delta}=4u^{[\alpha} \tilde {\mathcal X}(u)^{\beta][\gamma}u^{\delta]}\,,\quad  
\tilde {\mathcal X}(u)=\frac34 \frac{C_Q}{m}[S^2]^{\rm STF}\,,
\eeq
where $C_Q$ is a constant parameter. For neutron stars its value depends on the equation of state and varies roughly between 4 and 8 \cite{Laarakkers:1997hb}, whereas it is exactly $C_Q = 1$ for black holes \cite{Thorne:1980ru}. 

Therefore, the quadrupole tensor can be decomposed as
\beq
\label{Jdef}
J^{\alpha\beta\gamma\delta}=\frac34 \frac{C_Q}{m}  [J_{SS}-\frac13 s^2 J_{\perp}]^{\alpha\beta\gamma\delta}\,,
\eeq
where
\begin{eqnarray}
J_{SS}^{\alpha\beta\gamma\delta}&=& u^\alpha S(u)^\beta S(u)^\gamma u^\delta
-u^\alpha S(u)^\beta u^\gamma S(u)^\delta\nonumber\\
&&-S(u)^\alpha u^\beta S(u)^\gamma u^\delta+S(u)^\alpha u^\beta u^\gamma S(u)^\delta
\,,\nonumber\\ 
J_{\perp}^{\alpha\beta\gamma\delta}&=& u^\alpha P(u)^{\beta\gamma} u^\delta
-u^\alpha P(u)^{\beta\delta} u^\gamma\nonumber\\
&&-u^\beta P(u)^{\alpha \gamma} u^\delta
+u^\beta P(u)^{\alpha\delta} u^\gamma\,,
\end{eqnarray}
and in $J_{\perp}$ one can replace $P(u)$ by the metric $g$. 

The MPD equations \eqref{eq_MPD}--\eqref{TDconds} imply that the unit vectors $U$ and $u$ are related by
\beq
\label{reluUgen}
u^\mu=U^\mu+\frac1{m_0}D^{\mu \nu}_{\rm (quad)}U_\nu
+\frac1{m_0^2}S^{\mu\nu} F_{\rm (spin)}{}_{\nu}+O(S^3)\,,
\eeq 
where $m_0$ denotes the (conserved) bare mass of the extended body.
The spin-dependent effective mass $m$ is instead given by
\beq
m=m_0+m_J+O(S^3)\,,
\eeq
where
\beq
\label{mJdef}
m_J=\frac16J^{\alpha \beta \gamma \delta}R_{\alpha \beta \gamma \delta}\,.
\eeq

Finally, in stationary and axisymmetric spacetimes endowed with Killing symmetries there exist conserved quantities associated with the timelike Killing vector $\xi=\partial_t$ (the energy $E$) and the azimuthal Killing vector $\eta=\partial_\phi$ (the total angular momentum $J$) to all multipolar orders~\cite{ehlers77}, i.e., 
\begin{eqnarray}
\label{totalenergy}
E&=-\xi_\alpha P^\alpha +\frac12 S^{\alpha\beta}\nabla_\beta \xi_\alpha\,,\nonumber\\
J&=\eta_\alpha P^\alpha -\frac12 S^{\alpha\beta}\nabla_\beta \eta_\alpha\,,
\end{eqnarray}
respectively, where $\nabla_\beta \xi_\alpha=g_{t[\alpha,\beta]}$ and $\nabla_\beta \eta_\alpha=g_{\phi[\alpha,\beta]}$.

\section{Circular motion in a Schwarzschild spacetime}

Let us consider the Schwarzschild spacetime, with line element written in standard spherical-like coordinates $(t,r,\theta,\phi)$  given by
\beq 
\label{metric}
ds^2 = -fd t^2 + f^{-1} d r^2 
+ r^2 (d \theta^2 +\sin^2 \theta d \phi^2)\,,
\eeq
where $f=1-2M/r$. A natural orthonormal frame (adapted to the static observers, at rest with respect to the spatial coordinates) is the following 
\begin{eqnarray}
\label{frame}
e_{\hat t}&=&f^{-1/2}\partial_t\,, \qquad
e_{\hat r}=f^{1/2}\partial_r
\,, \nonumber\\
e_{\hat \theta}&=&\frac{1}{r}\partial_\theta\,, \qquad
e_{\hat \phi}=\frac{1}{r\sin \theta}\partial_\phi\,,
\end{eqnarray}
where $\{\partial_t, \partial_r, \partial_\theta, \partial_\phi\}$ is the coordinate frame.
The orthonormal component along $-\partial_\theta$ which is perpendicular to the equatorial plane will be referred to as ``along the positive $z$-axis," and will be denoted by the index $\hat z$, so that $e_{\hat z}=-e_{\hat \theta}$. 

It is convenient to decompose the spin vector $S(u)$ in magnitude ($s$) and direction ($N(u)$): $S(u)=s N(u)$, with $N(u)$ unitary, spacelike and orthogonal to $u$, namely $u\cdot N(u)=0$,  $N(u)\cdot N(u)=1$. 
The spin-induced quadrupole tensor \eqref{Jdef} thus reads 
\beq
J^{\alpha\beta\gamma\delta}=\frac34 \frac{C_Q}{m} s^2 [J_{NN}-\frac13 J_{\perp}]^{\alpha\beta\gamma\delta}\,,
\eeq
since $J_{SS}=s^2 J_{NN}$.
Let us assume that $N(u)$ be aligned with the $z$-axis of an orthonormal frame adapted to $u=e_0$: $N(u)=e_{\hat z}=-e_{\hat \theta}$, so that 
\beq
\label{J_expr}
J^{\alpha\beta\gamma\delta}=\frac34 \frac{C_Q}{m}s^2 \left[(e_0\wedge e_{\hat \theta})^{\alpha\beta}(e_0\wedge e_{\hat \theta})^{\gamma\delta}-\frac13 J_{\perp}^{\alpha\beta\gamma\delta}
 \right]\,.
\eeq
To make this expression more compact we can introduce an orthonormal frame adapted to $u$, $\{e_\alpha \}$, with $e_0=u$ and $e_2=e_{\hat \theta}$ and $e_1$ and $e_3$ spanning the $\theta=$const. hyperplane. By using this frame one finds the following representation for $P(u)$:
\beq
P(u)=e_1\otimes e_1+e_2\otimes e_2+e_3\otimes e_3\,,
\eeq
and hence one can replace the various terms in \eqref{J_expr} with tensor products of frame vectors. For example,
\begin{eqnarray}
u^\alpha P(u)^{\beta\gamma} u^\delta &=& e_0^\alpha (e_1^\beta e_1^\gamma +e_2^\beta e_2^\gamma+e_3^\beta e_3^\gamma) e_0^\delta\nonumber\\
&=&[E_{0110}+E_{0220}+E_{0330}]^{\alpha\beta\gamma\delta}\nonumber\\
u^\alpha P(u)^{\beta\delta} u^\gamma &=& [E_{0101}+E_{0202}+E_{0303}]^{\alpha\beta\gamma\delta}\nonumber\\
u^\beta P(u)^{\alpha\gamma} u^\delta &=& [E_{1010}+E_{2020}+E_{3030}]^{\alpha\beta\gamma\delta}\nonumber\\
u^\beta P(u)^{\alpha\delta} u^\gamma &=& [E_{1001}+E_{2002}+E_{3003}]^{\alpha\beta\gamma\delta}\,,
\end{eqnarray}
where we have adopted the multi-tensor product notation
\beq
E_{0112}^{\alpha\beta\gamma\delta}
=e_0^\alpha e_1^\beta e_1^\gamma e_0^\delta\,,
\eeq
etc.

The final expression of the quadrupole tensor is the following
\beq
\label{final_J}
J^{\alpha\beta\gamma\delta}={\mathcal J}\left( e_{01}^{\alpha\beta} e_{01}^{\gamma\delta}-2  e_{02}^{\alpha\beta} e_{02}^{\gamma\delta} +e_{03}^{\alpha\beta} e_{03}^{\gamma\delta}\right)\,,   
\eeq
where 
\beq
{\mathcal J}=\frac14 \frac{C_Q}{m}s^2\,,
\eeq
and we have used the wedge-product notation
\beq
 e_{02}^{\alpha\beta}=e_0^\alpha e_2^\beta-e_0^\beta e_2^\alpha\,,
\eeq
etc.
The compact and elegant expression \eqref{final_J} for the quadrupole tensor makes trivial any tensor contraction. 
For instance, 
the quadrupole correction \eqref{mJdef} to the mass of the body turns out to be
\begin{eqnarray}
\label{mJdef2}
m_J
&=&\frac23{\mathcal J}[R_{0101}-2R_{0202}+R_{0303}]\nonumber\\
&=&\frac23{\mathcal J}[E(u)_{11}-2E(u)_{22}+E(u)_{33}]\nonumber\\
&=& -2{\mathcal J}E(u)_{22}\,,
\end{eqnarray}
where we have introduced the electric part of the Riemann with respect to $u$, $E(u)_{\alpha\beta}=R_{0\alpha0\beta}$ (symmetric and tracefree).

\subsection{Circular orbits}

Let the body with spin vector $S(u)=s\,e_{\hat z}$ and spin-induced quadrupole tensor \eqref{final_J} move along a circular orbit on the equatorial plane, with unit tangent vector $U$ parametrized either by the (constant) angular velocity $\zeta$ with respect to infinity or, equivalently, by the (constant) linear velocity $\nu$ with respect to the static observers as
\beq
U=\Gamma [\partial_t +\zeta \partial_\phi ]
=\gamma [e_{\hat t} +\nu e_{\hat \phi}]\,, \qquad 
\nu = rf^{-1/2}\zeta\,,
\eeq
with normalization factors
\beq
-\Gamma^{-2}=g_{tt}+\zeta^2 g_{\phi\phi}\,, \qquad 
\gamma=(1-\nu^2)^{-1/2}=\Gamma f^{1/2}\,.
\eeq
The parametric equations of the orbit are then given by
\beq
t=t_0+\Gamma \tau\,,\quad 
r=r_0\,,\quad 
\theta=\frac{\pi}{2}\,,\quad
\phi=\phi_0+\Gamma\zeta\tau\,.
\eeq
The MPD equations imply that the direction $u$ of the 4-momentum is also tangent to a circular orbit with different angular and linear velocities $\zeta_u$ and $\nu_u$, i.e.,
\beq
u=\Gamma_u [\partial_t +\zeta_u \partial_\phi ]
=\gamma_u [e_{\hat t} +\nu_u e_{\hat \phi}]\,,
\eeq
with similar relation for the corresponding normalization factors.
An orthonormal frame adapted to $e_0=u$ is then built with the spatial triad
\begin{eqnarray}
e_1&=&e_{\hat r}\,, \qquad
e_2=e_{\hat \theta}\,, \nonumber\\
e_3&=&\bar\Gamma_u [\partial_t +\bar\zeta_u \partial_\phi ]
=\gamma_u [\nu_ue_{\hat t} + e_{\hat \phi}]\,,
\end{eqnarray}
with $\bar\Gamma_u=\Gamma_u\nu_u$ and $\bar\zeta_u=f^{1/2}/(r\nu_u)$.
Therefore, the quadrupole correction \eqref{mJdef2} to the mass of the body turns out to be
\beq
\label{mJdef3}
m_J=-2{\mathcal J}\frac{M}{r^3}\gamma_u^2(1+2\nu_u^2)\,.
\eeq

Under the assumptions of equatorial motion and spin vector aligned with the $z$-axis the MPD equations imply that the signed spin magnitude $s$ is a constant of
motion (see e.g., \cite{Bini:2015zya}).
Therefore, we introduce the dimensionless spin parameter
\beq
{\hat s}=\frac{s}{m_0M}\,,
\eeq
which we will take as a smallness indicator. 
Hereafter, all spin-dependent quantities are then understood to be evaluated up to the order $O(\hat{s}^2)$.

\subsubsection{Frequencies $\zeta$ and $\zeta_u$}

The solutions for the frequencies $\zeta$ and $\zeta_u$ are
\begin{eqnarray}
M\zeta &=& u_0^{3/2}\left[1-\frac32 u_0^{3/2}\hat s\right.\nonumber\\
&+&\left.
\frac34u_0^2\left(\frac72u_0+ C_Q(1-2u_0)\right)\hat s^2\right]
+O(\hat s^3)
\,,\nonumber\\
M\zeta_u &=& u_0^{3/2}\left[1-\frac32 u_0^{3/2}\hat s\right.\nonumber\\
&+&\left.
\frac34u_0^2\left(-\frac12u_0+ C_Q(1+2u_0)\right)\hat s^2\right]
+O(\hat s^3)
\,,\nonumber\\ 
\end{eqnarray}
so that $M(\zeta_u -\zeta)=3(C_Q-1)u_0^{9/2}\hat s^2+O(\hat s^3)$, where we have used the dimensionless (inverse) radial variable
\beq
u_0
=\frac{M}{r_0}\,.
\eeq
Both $M\zeta$ and $M\zeta_u$ correspond to spin and spin-square modifications of the circular geodesic (Keplerian) value
$M\zeta_K=u_0^{3/2}$.

It is useful to introduce the dimensionless frequency variable $y=(M\zeta)^{2/3}$, which to second order in spin reads
\begin{eqnarray}
y &=& u_0\left(1-u_0^{3/2}\hat s+\frac12 u_0^3 \hat s^2\right)\nonumber\\
&+&
\frac{1}{2}u_0^3\left[ 1+(C_Q-1)(1-2u_0) \right]\hat s^2
+O(\hat s^3)
\,,\nonumber\\
\end{eqnarray}
with inverse
\begin{eqnarray}
\label{u0vsy0}
u_0 &=& y\left(1+y^{3/2}\hat s\right.\nonumber\\
&-&\left.
\frac12 y^2[1-4y+(C_Q-1)(1-2y)] \hat s^2\right)
+O(\hat s^3)
\,.\nonumber\\
\end{eqnarray}

\subsubsection{Normalization factors $\Gamma_0$ and $\Gamma_u$}

The normalization factors $\Gamma$ and $\Gamma_u$ are given by
\begin{eqnarray}
\Gamma &=& \frac{1}{\sqrt{1-3u_0}} -\frac{3u_0^{5/2}}{2(1-3u_0)^{3/2}}\hat s\nonumber\\
&+&\frac{3u_0^3}{4(1-3u_0)^{3/2}}\left[ C_Q(1-2 u_0)\right.\nonumber\\
&+&\left.
 \frac12u_0\frac{10-21u_0}{1-3u_0}\right]\hat s^2
+O(\hat s^3)
\,,\nonumber\\
\Gamma_u &=& \frac{1}{\sqrt{1-3u_0}} -\frac{3u_0^{5/2}}{2(1-3u_0)^{3/2}}\hat s\nonumber\\
&+&\frac{3u_0^3}{4(1-3u_0)^{3/2}}\left[ C_Q(1+2 u_0)\right.\nonumber\\
&+&\left.
 \frac12u_0\frac{2+3u_0}{1-3u_0}\right]\hat s^2
+O(\hat s^3)
\,.\nonumber\\
\end{eqnarray}
so that $\Gamma-\Gamma_u=-3(C_Q-1)u_0^4\hat s^2/(1-3u_0)^{3/2}+O(\hat s^3)$.

The redshift variable $z_1^{(0)}=\Gamma^{-1}$ as a function of $y$ is then given by
\begin{eqnarray}
\label{z10}
z_1^{(0)}(y)&=&\sqrt{1-3y}-\frac{3y^4}{2\sqrt{1-3y}}\hat s^2
+O(\hat s^3)
\nonumber\\
&=&1-\frac32y-\frac98y^2-\frac{27}{16}y^3-\frac{405}{128}y^4\nonumber\\
&-&\frac{1701}{256}y^5-\frac{15309}{1024}y^6+O(y^7)\nonumber\\
&+&
\left(-\frac32y^4-\frac94y^5-\frac{81}{16}y^6+O(y^7)\right)\hat s^2
+O(\hat s^3)\,.\nonumber\\
\end{eqnarray}

Furthermore, the quadrupole correction \eqref{mJdef3} to the mass of the body turns out to be
\beq
m_J=- m_0 C_Q  \frac{u_0^3}{2(1-3u_0)}\hat s^2\,.
\eeq

\subsubsection{Conserved energy and angular momentum}

The conserved energy and angular momentum \eqref{totalenergy} as functions of the frequency variable $y$ are given by
\begin{eqnarray}
\label{EJspin}
\frac{E}{m_0}&=& \frac{1-2y}{\sqrt{1-3y}} -\frac{y^{5/2}}{ (1-3 y)^{1/2}}\hat s\nonumber\\
&+&\frac{ y^3 [(1-3y)(1-4y)+(C_Q-1)(1-2y)]}{2(1-3 y)^{3/2}}\hat s^2\nonumber\\
&+&O(\hat s^3)
\,,\nonumber\\
\frac{J}{m_0M} &=&  \frac{1}{\sqrt{y(1-3y)}}+\frac{1-4y}{\sqrt{1-3y}}\hat s\nonumber\\
&+&\frac{y^{3/2}[(1-3y)(2-7y)+(C_Q-1)(2-5y)]}{2(1-3 y)^{3/2}}\hat s^2\nonumber\\
&+&O(\hat s^3)\,.
\end{eqnarray}

\section{Energy momentum tensor}

Following Ref. \cite{Steinhoff:2009tk}, the energy momentum tensor of a quadrupolar particle is given by
\beq
T^{\alpha\beta}=\int d\tau \frac{1}{\sqrt{-g}}{\mathcal T}^{\alpha\beta}\,,
\eeq
where
\begin{eqnarray}
{\mathcal T}^{\alpha\beta}&=&
\left(U^{(\alpha}P^{\beta)}+\frac13R_{\gamma\delta\epsilon}{}^{(\alpha}J^{\beta)\epsilon\delta\gamma}\right)\delta^{(4)}
\nonumber\\
&&-\nabla_\gamma\left(S^{\gamma(\alpha}U^{\beta)}\delta^{(4)}\right)\nonumber\\
&& -\frac23\nabla_\delta\nabla_\gamma\left(J^{\delta(\alpha\beta)\gamma}\delta^{(4)}\right)\,.
\end{eqnarray}
Here $\delta^{(4)}$ denotes the 4-dimensional delta function centered on the particle's worldline, i.e.,
\begin{eqnarray}
\delta^{(4)}&\equiv& \delta^{(4)}(x^\alpha -x^\alpha(\tau))\nonumber\\
&=&\delta (t-\Gamma \tau)\delta^{(3)}(x^a -x^a(\tau))\nonumber\\
&=&\frac1{\Gamma}\delta \left(\tau-\frac{t}{\Gamma}\right)\delta^{(3)}(x^a -x^a(t))\,,
\end{eqnarray}
where
\beq
\delta^{(3)}(x^a -x^a(t))=\delta(r-r_0)\delta (\theta-\pi/2) \delta (\phi-\zeta t)
\equiv\delta^{(3)}\,.
\eeq
Integration over $\tau$ then yields
\begin{eqnarray}
T^{\alpha\beta}&=&
\frac{1}{\sqrt{-g}}\frac{1}{\Gamma}\left(m U^{(\alpha}u^{\beta)}+\frac13R_{\gamma\delta\epsilon}{}^{(\alpha}J^{\beta)\epsilon\delta\gamma}\right)\delta^{(3)}\nonumber\\
&&-\frac{1}{\sqrt{-g}}\nabla_\gamma\left(\frac{1}{\Gamma}S^{\gamma(\alpha}U^{\beta)}\delta^{(3)}\right)\nonumber\\
&& -\frac23\frac{1}{\sqrt{-g}} \nabla_\delta\nabla_\gamma\left(\frac{1}{\Gamma} J^{\delta(\alpha\beta)\gamma}\delta^{(3)}\right)\,.
\end{eqnarray}
 
The energy momentum tensor thus results in the sum of three pieces
\beq
T_{\mu\nu}=T^{\hat s^0}_{\mu\nu}+\hat sT^{\hat s^1}_{\mu\nu}+\hat s^2T^{\hat s^2}_{\mu\nu}\,,
\eeq
which are listed below.
We use the following notation for the first derivatives of delta functions
\begin{eqnarray}
\label{deltas}
\delta_{r}^{(3)}&=&\delta{}'(r-r_0)\delta (\theta-\pi/2) \delta (\phi-\zeta t)\,, \nonumber\\
\delta_{\theta}^{(3)}&=&\delta(r-r_0)\delta' (\theta-\pi/2) \delta (\phi-\zeta t)\,, \nonumber\\
\delta_{\phi}^{(3)}&=&\delta(r-r_0)\delta (\theta-\pi/2)\delta{}'(\phi-\zeta t)\,,
\end{eqnarray}
and similarly for the second derivatives.

\begin{enumerate}
  \item The $\hat s^0$ term is given by
\beq
T^{\hat s^0}_{\mu\nu}=\frac{m_0u_0}{\sqrt{1-3u_0}} \delta^{(3)} 
\begin{pmatrix}
\frac{(1-2 u_0)^2 u_0}{M^2}& 0& 0& -\frac{(1-2 u_0)  u_0^{1/2}}{M}\cr
0& 0& 0& 0\cr
0& 0& 0& 0\cr
{\rm sym}& 0& 0&  1\cr 
\end{pmatrix}\,.
\eeq
Here, we identify  the tensor  which is multiplied by $\delta^{(3)}$, through the prefactor $\frac{m_0u_0}{\sqrt{1-3u_0}}$ and its nonzero components: $tt$, $t\phi$ and $\phi\phi$. This notation  is used in Table  \ref{X_r_mu_nu}.

\item The $\hat s^1$ term is given by
\beq
T^{\hat s}_{\mu\nu}=X_{\mu\nu}\delta^{(3)} 
+X_r{}_{\mu\nu}\delta^{(3)}_{r}
+X_\phi{}_{\mu\nu}\delta^{(3)}_{\phi}\,,
\eeq
where the nonvanishing components of the tensors $X_{\mu\nu}$, $X_r{}_{\mu\nu}$ and $X_\phi{}_{\mu\nu}$ are listed in Table 
\ref{X_r_mu_nu}.

\item  The $\hat s^2$ term is given by 
\beq
T^{\hat s^2}_{\mu\nu}=T^{\hat s^2 C_Q^0}_{\mu\nu}+C_Q T^{\hat s^2 C_Q^1}_{\mu\nu}\,,
\eeq
with
\beq
T^{\hat s^2 C_Q^0}_{\mu\nu}=Y_{\mu\nu}\delta^{(3)}
+Y_{r}{}_{\mu\nu}\delta^{(3)}_{r}
+Y_{\phi}{}_{\mu\nu}\delta^{(3)}_{\phi}\,,
\eeq
and
\begin{eqnarray}
T^{\hat s^2 C_Q^1}_{\mu\nu}&=&
Z_{\mu\nu}\delta^{(3)}
+Z_{r}{}_{\mu\nu}\delta^{(3)}_{r}
+Z_{\phi}{}_{\mu\nu}\delta^{(3)}_{\phi}\nonumber\\
&+&
Z_{rr}{}_{\mu\nu}\delta^{(3)}_{rr}
+Z_{\theta\theta}{}_{\mu\nu}\delta^{(3)}_{\theta\theta}
+Z_{\phi\phi}{}_{\mu\nu}\delta^{(3)}_{\phi\phi}
\,.\nonumber\\
\end{eqnarray}
The nonvanishing components of the various tensors are listed in Table \ref{X_r_mu_nu}.

\end{enumerate}

% table 1

\begin{table*}
\caption{\label{X_r_mu_nu} List of nonvanishing components (modulo symmetries) of the various symmetric source tensors}
\begin{ruledtabular}
\begin{tabular}{lllll}
Tensor & prefactor & component & component & component \\
\hline
$X_{\mu\nu}$ & $m_0 \frac{u_0^{3/2}}{(1-3 u_0)^{3/2}}$ & $tt$ & $t\phi$ & $\phi\phi$\\
& &$-\frac{(1-2 u_0) (36 u_0^2-23 u_0+4) u_0^2}{2 M^2}$ & $\frac{ (18 u_0^2-14 u_0+3) u_0^{3/2}}{2 M}$ & $\frac{(27 u_0^2-16 u_0+2)}{2}$\\
&& $rr$ && \\
&& $-\frac{(1-3 u_0)^2 u_0^2}{ M^2 (1-2 u_0)}$ && \\
\hline
$X_r{}_{\mu\nu}$ & $-m_0\frac{u_0^{1/2}(1-2u_0)}{ (1-3 u_0)^{1/2}} $ & $tt$ & $t\phi$ & $\phi\phi$\\ 
& &$\frac{(1-2 u_0)  u_0^2}{ M}$ & $-\frac{(1-u_0) u_0^{1/2}}{2}$ &$ M $\\
\hline
$X_\phi{}_{\mu\nu}$ & $m_0 \frac{ u_0^{5/2} (1-3 u_0)^{1/2}}{2 M (1-2 u_0)}$ & $tr$ & $r\phi$ & \\
& & $-\frac{ u_0^{1/2}(1-2u_0)}{M}$ & $1$ & \\
\hline
\hline
$Y_{\mu\nu}$ & $m_0\frac{3u_0^3}{2(1-3u_0)^{1/2}}$ & $tt$ & $t\phi$ & $\phi\phi$\\
&& $\frac{ (1-2 u_0)^2 u_0^2 (87 u_0^2-50 u_0+8)}{4 M^2 (1-3 u_0)^{2}}$ & $\frac{u_0^{3/2} (56 u_0-160 u_0^2-7+156 u_0^3)}{4(1-3 u_0)^{2} M}$ & $ \frac{(123 u_0^3-122 u_0^2+40 u_0-4) }{4(1-3 u_0)^{2}}$\\
\hline
$Y_{r}{}_{\mu\nu}$ &$m_0 \frac{u_0^2(1-2u_0)}{2(1-3u_0)^{3/2}}$ & $tt$ & $t\phi$ & $\phi\phi$ \\
&&$\frac{3 u_0^2 (1-2 u_0)^2}{M}$ & $\frac{3 (7 u_0-3) u_0^{3/2}}{2}$ & $3 M (1-2 u_0)$\\
\hline
$Y_{\phi}{}_{\mu\nu}$&$m_0\frac{3 u_0^4}{4M(1-3 u_0)^{1/2}} $ & $tr$ & $r\phi$ &\\
&& $ -\frac{u_0^{3/2}}{M}$ & $-\frac{1-4 u_0}{1-2 u_0} $ & \\
\hline
\hline
$Z_{rr}{}_{\mu\nu}$ & $m_0\frac{ u_0 (1-2u_0)}{6(1-3 u_0)^{1/2}}$ & $tt$ & $t\phi$ & $\phi\phi$ \\
&& $u_0 (1-2 u_0)^2 $ &$ -u_0^{1/2} (1-2 u_0) M$ & $M^2$\\
\hline
$Z_{\theta\theta}{}_{\mu\nu}$ & $-m_0\frac{ u_0^3}{3(1-3 u_0)^{1/2}}$ & $tt$ & $t\phi$ & $\phi\phi$\\
&& $\frac{ u_0 (1-2 u_0)^2}{M^2} $ & $-\frac{(1-2 u_0)  u_0^{1/2}}{M} $  &$1$ \\
\hline
$Z_{\phi\phi}{}_{\mu\nu}$ & $m_0\frac{ u_0^3(1-3 u_0)^{1/2}}{6}$ & $tt$ & $t\phi$ & $\phi\phi$\\
&&  $ \frac{u_0(1-2u_0)}{M^2} $ & $-\frac{u_0^{1/2}}{M}$ & $\frac{1}{1-2 u_0}$\\
\hline
$Z_{r}{}_{\mu\nu}$ & $m_0\frac{ u_0^2}{6(1-3 u_0)^{1/2}}$& $tt$ & $t\phi$ & $\phi\phi$\\
&& $-\frac{(16 u_0-5)  u_0 (1-2 u_0)^2}{M}$ & $(1-2 u_0) (10 u_0-3) u_0^{1/2}$ & $M (1-4 u_0)$\\
\hline
$Z_{\phi}{}_{\mu\nu}$ &$m_0\frac{ u_0^4 (1-3 u_0)^{1/2}}{3M} $ & $tr$ & $r\phi$ & \\
&& $\frac{ u_0^{1/2} }{M}$ & $-\frac{1}{1-2 u_0} $ & \\
\hline
$Z_{\mu\nu}$ & $m_0\frac{ u_0^3}{3 (1-3 u_0)^{1/2}}$& $tt$ & $t\phi$ & $\phi\phi$\\
&& $-\frac{(468 u_0^3-432 u_0^2+125 u_0-12) u_0 (1-2 u_0)}{4(1-3 u_0) M^2} $ & $-\frac{ (72 u_0^2-54 u_0+13) (1-2 u_0) u_0^{1/2}}{4 M (1-3 u_0)} $ & $\frac{18 u_0^2-63 u_0+22}{4(1-3 u_0)}$\\
&& $rr$ & $\theta\theta$ & \\
&& $-\frac{ u_0^2}{ M^2(1-2 u_0)}$ & $-2$ & \\
\end{tabular}
\end{ruledtabular}
\end{table*}

\section{First-order metric perturbations and Detweiler's redshift invariant $z_1$}

Let us consider now an extended body as above still moving along an accelerated equatorial circular orbit with spin vector aligned with the $z$-axis according to the MPD model, but in a perturbed Schwarzschild spacetime.
We are interested in computing Detweiler's redshift invariant
\beq
\label{z1_def_s_quad}
z_1=\sqrt{1-2 u_0-\frac{y^3}{u_0^2}-qh^{\rm R}_{kk}}\,,
\eeq
to first order in the mass ratio $q\equiv m_0/M\ll1$, where $h^{\rm R}_{kk}=h^{\rm R}_{\alpha\beta}k^\alpha k^\beta$ is the regularized value of the double contraction of the metric perturbation $h_{\alpha\beta}(x^\mu)$ induced by the extended body with the helical Killing vector $k=\partial_t+\zeta\partial_\phi$.
Therefore, we need $h^{\rm R}_{kk}$ to second order in spin 
\beq
h^{\rm R}_{kk}=h^{\rm R}_{kk\,(0)}(y)+\hat s h^{\rm R}_{kk\,\hat s}(y)+\hat s^2 h^{\rm R}_{kk\,\hat s^2}(y)\,,
\eeq
(hereafter, we remove the label R for simplicity) as well as the perturbed relation between the variables $u_0$ and $y$
\begin{eqnarray}
\label{u_rel}
u_0 &=& y \left[1+y^{3/2}\hat s-\frac12 y^2 [1-4y+(C_Q-1)(1-2y)]\hat s^2\right] \nonumber\\
&&
+q[f_0(y)+\hat s f_{\hat s}(y)+\hat s^2 f_{\hat s^2}(y)]\,,
\end{eqnarray}
where the functions $f_0(y)$, $f_{\hat s}(y)$ and $f_{\hat s^2}(y)$ are determined by solving the MPD equations in the perturbed Schwarzschild spacetime.
Inserting these relations in Eq. \eqref{z1_def_s_quad} and expanding to first order in $q$ and to second order in $\hat s$ then gives
\beq
z_1(y)=z_1^{(0)}(y)+q \left(z_1^{(1)\hat s^0}(y)+\hat s z_1^{(1)\hat s^1}(y)+\hat s^2 z_1^{(1)\hat s^2}(y)\right)\,,
\eeq
where the unperturbed value $z_1^{(0)}(y)$ is given by Eq. \eqref{z10} and the first-order self-force (1SF) contributions are
\begin{eqnarray}
z_1^{(1)\hat s^0}(y)&=& -\frac{1}{2\sqrt{1-3y}} h_{kk\,(0)}(y)
\,, \nonumber\\
z_1^{(1)\hat s^1}(y)&=& -\frac{1}{2\sqrt{1-3y}} \left[6y^{3/2} f_0(y)+h_{kk\,\hat s}(y)\right]
\,, \nonumber\\
z_1^{(1)\hat s^2}(y)&=& -\frac{1}{2\sqrt{1-3y}} \left[
\frac{3y^4}{2(1-3y)}  h_{kk\,(0)}(y)\right.\nonumber\\
&& 
-3[(1-2 y)(C_Q-1)+1] y^2 f_0(y) \nonumber\\
&& \left. 
+6 y^{3/2}f_{\hat s}(y) + h_{kk\,\hat s^2}(y)\right]\,.
\end{eqnarray}
Therefore, only the unknown functions $f_0(y)$ and $f_{\hat s}(y)$ entering the relation \eqref{u_rel} between $u_0$ and $y$ are needed.
They have been already determined in Ref. \cite{Bini:2018zde}  
\begin{eqnarray}
f_0(y)&=& \frac1{6y}M[\partial_r h_{kk\,(0)}](y)\,,\nonumber\\
f_{\hat s}(y)&=& 
2y^{3/2}f_0(y)-\frac{2}{3y^{1/2}} M\Omega_{1{\hat s}}(y)
\,,
\end{eqnarray}
where $M\Omega_{1{\hat s}}=u_0^{3/2}\tilde\Omega_{1{\hat s}}(u_0)$ evaluated at $u_0=y$, with $\tilde\Omega_{1{\hat s}}$ given by Eq. (B11) of Ref. \cite{Bini:2018zde}, i.e.,

\begin{widetext}
\begin{eqnarray}
\tilde\Omega_{1{\hat s}}&=&
-\frac{u_0^{3/2}}{4(1-2u_0)^2}h_{kk}^{(0)}
+\frac{(5-12u_0)u_0^{3/2}}{4}h_{rr}^{(0)}
-\frac{u_0^2(3-4u_0)(1-3u_0)}{2M(1-2u_0)^2}h_{t\phi}^{(0)}
-\frac{(1-3u_0)(2-5u_0+4u_0^2)u_0^{5/2}}{4M^2(1-2u_0)^2}h_{\phi\phi}^{(0)}\nonumber\\
&&
-\frac{M^2u_0^{-3/2}}{4}[\partial_{rr} h_{kk}^{(0)}]_{r=M/u_0}
-\frac{M}{4u_0}(1-3u_0)[\partial_{rr} h_{\phi k}^{(0)}]_{r=M/u_0}
+\frac{M}{4u_0}(1-3u_0)[\partial_{r\bar\phi} h_{rk}^{(0)}]_{r=M/u_0}\nonumber\\
&&
-\frac14(1-3u_0)[\partial_{\bar\phi} h_{rk}^{(0)}]_{r=M/u_0}
+\frac{Mu_0^{1/2}}{4(1-2u_0)}[\partial_r h_{kk}^{(0)}]_{r=M/u_0}
-\frac{M}{4u_0^2}[\partial_r h_{kk}^{(1)}]_{r=M/u_0}\nonumber\\
&&
+\frac{M}{4}(1-2u_0)(1-3u_0)u_0^{-1/2}[\partial_r h_{rr}^{(0)}]_{r=M/u_0}
+\frac{(1-3u_0)}{4(1-2u_0)}[\partial_r h_{t\phi}^{(0)}]_{r=M/u_0}\nonumber\\
&&
+\frac{(1-3u_0)(2-3u_0)u_0^{3/2}}{4M(1-2u_0)}[\partial_r h_{\phi\phi}^{(0)}]_{r=M/u_0}
\,.
\end{eqnarray}
\end{widetext}

The second order in spin 1SF contribution to the redshift finally reads
\begin{eqnarray}
\label{z1squadfin}
z_1^{(1)\hat s^2}(y)
&=& -\frac{3y^4}{4(1-3y)^{3/2}}  h_{kk\,(0)}(y)\nonumber\\  
&+& \frac{y [1-4 y+(1-2 y) (C_Q-1)]}{4\sqrt{1-3y}}M[\partial_r h_{kk\,(0)}](y)\nonumber\\
&+&\frac{2y^{5/2}}{\sqrt{1-3y}}\tilde\Omega_{1{\hat s}}(y)-\frac{1}{2\sqrt{1-3y}} h_{kk\,\hat s^2}(y)\,.
\end{eqnarray}
Its determination requires the separate GSF computations of $ h_{kk\,(0)}$, $\partial_r h_{kk\,(0)}$, $\tilde\Omega_{1{\hat s}}$ and $h_{kk\,\hat s^2}$.
Each of such terms is gauge-dependent and only their combination \eqref{z1squadfin} leads to the gauge-invariant quantity $z_1^{(1)\hat s^2}$.
We use here the Teukolsky formalism in a radiation gauge and the related CCK procedure to reconstruct the radiative part of the metric perturbation. 
This method is well established in the literature (see, e.g., Ref. \cite{Shah:2012gu}), so we will skip all unnecessary details.
The nonradiative part of the metric is, instead, evaluated by using the RWZ approach as in our previous work \cite{Bini:2018zde}.

We refer to that work and references therein for a detailed account of the non-spinning terms.
Ref. \cite{Bini:2018zde} also contains the necessary information to determine the linear-in-spin 1SF correction to the frequency $\tilde\Omega_{1{\hat s}}$. 
Therefore, we will provide below some details on the computation of the quadratic-in-spin term $h_{kk\,\hat s^2}$.

\subsection{Computing $h_{kk\,\hat s^2}$}

The radiation-gauge metric perturbation approach gives a PN expansion of the radiative $\ell$-modes ($\ell\geq 2$), $h_{kk}^{\ell\,\rm(rad)}$, of the retarded value of $h_{kk}$. These PN-type solutions provides information on the large-$\ell$ behavior of the modes, and should be combined with MST-type solutions (for certain low values of $\ell=2,3,\ldots$) in order to reach a high-PN level of accuracy of the final result.
The non-radiative part of the perturbation ($\ell=0,1$) must be computed separately, and corresponds to mass and angular momentum perturbations of the background, up to gauge modes.
The relevant components of the exterior $(+)$ and interior $(-)$ metric perturbations (evaluated at $\theta=\pi/2$) are found to be
\beq
qh_{tt\,(+)}^{\rm(nonrad)}=\frac{2\delta M}{r}\,, \qquad
qh_{t\phi\,(+)}^{\rm(nonrad)}=-\frac{2\delta J}{r}\,,
\eeq
and
\begin{eqnarray}
qh_{tt\,(-)}^{\rm(nonrad)}&=&
\frac{2\delta Mu_0f}{Mf_0}\left[1-\frac{u_0^{3/2}}{f_0}(2-3u_0)\hat s(1+{\mathcal B}_0\hat s)\right]
\,, \nonumber\\
qh_{t\phi\,(-)}^{\rm(nonrad)}&=&
-\frac{2\delta Ju_0^3r^2}{M^3}\left[1-\frac32u_0^{1/2}(1-u_0)\hat s(1+{\mathcal D}_0\hat s)\right]
\,, \nonumber\\
\end{eqnarray}
respectively (see Appendix A for details).
Here $\delta M \equiv E $ and $\delta J \equiv  J $ are given by the conserved energy and angular momentum \eqref{EJspin} of the extended body, respectively, whereas the coefficients ${\mathcal B}_0$ and ${\mathcal D}_0$ are given by Eqs. \eqref{B0def} and \eqref{D0def}, respectively.

The full retarded solution is then
\beq
h_{kk}=\sum_{\ell=2}^\infty h_{kk}^{\ell\,\rm(rad)}+h_{kk}^{\rm(nonrad)}
=\sum_{\ell=0}^\infty h_{kk}^\ell\,,
\eeq
which needs to be suitably regularized, being divergent at the location of the source.
This is done standardly by removing the divergent large-$\ell$ behavior of the radiative modes as well as by taking the average between the two radial limits $r\to r_0^-$ (left) and $r\to r_0^+$ (right), leading to the following regularized value $h_{kk}^{\rm R}$ of $h_{kk}$
\beq \label{reghkk}
h_{kk}^{\rm R}=\sum_\ell\left[\langle h_{kk}^\ell \rangle-B(y;\ell)\right]\,,
\eeq
where 
\beq
\langle h_{kk}^\ell \rangle = \frac12(h_{kk\,(+)}^\ell+h_{kk\,(-)}^\ell)\,,
\eeq
and the \lq\lq subtraction term'' $B(y;\ell)$ is of the type
\beq
\label{Btermgen}
B(y;\ell)=\ell(\ell+1)b_0(y)+b_1(y)\,,
\eeq
with $b_0(y)=O(\hat s^2)$.
The subtraction terms for $h_{kk\,(0)}$ and $h_{kk\,\hat s}$ are given by Eqs. (5.12)--(5.13) of Ref. \cite{Bini:2018zde}, where a slight different notation is used ($B_{(0)}$ and $B_{\hat s}$ stand for $b_{1(0)}(y)$ and $b_{1\hat s}(y)$, respectively).
The subtraction term for $h_{kk\,\hat s^2}$ is given by Eq. \eqref{Btermgen} above with
\begin{eqnarray}
b_{0\hat s^2}(y)&=&
\frac12 C_Q y^3-\frac{27}{16} C_Q y^4+\frac{15}{128} C_Q y^5+\frac{545}{2048} C_Q y^6\nonumber\\
&+&
\frac{19965}{32768} C_Q y^7+\frac{368847}{262144} C_Q y^8+\frac{6875451}{2097152} C_Q y^9\nonumber\\
&+&
O(y^{10})
\,, \nonumber\\
b_{1\hat s^2}(y)&=&
-\frac12 C_Q y^3+\frac{51}{64} C_Q y^4+\left(\frac{23}{4}-\frac{765}{256} C_Q\right) y^5\nonumber\\
&+&\left(\frac{6205}{8192} C_Q+\frac98\right) y^6+\left(\frac{31659}{16384} C_Q+\frac{1131}{512}\right) y^7\nonumber\\
&+&\left(\frac{2657}{512}+\frac{5229453}{1048576} C_Q\right) y^8\nonumber\\
&+&
\left(\frac{54578889}{4194304} C_Q+\frac{938349}{65536}\right) y^9
+O(y^{10})
\,, \nonumber\\
\end{eqnarray}

The final result for the regularized value of the quadratic-in-spin term $h_{kk\,\hat s^2}^{\rm R}$ (including the MST solutions for $l=2,3,4$) is given in Table \ref{hkklist} together with the corresponding expressions for the other quantities needed to compute the redshift invariant (which require the MST solutions up to $l=6$).

% table 2

\begin{table*}
\caption{\label{hkklist}  
List of the regularized values of the various GSF quantities need to compute the redshift invariant.}
\begin{ruledtabular}
\begin{tabular}{ll}
GSF quantity & PN expansion \\
\hline
$h_{kk\,(0)}^{\rm R}$ & $-2y+5 y^2+\frac54 y^3 +\left(-\frac{1261}{24}+\frac{41}{16}\pi^2\right) y^4 +\left(\frac{157859}{960}-\frac{2275}{256}\pi^2-\frac{256}{5}\gamma-\frac{512}{5}\ln(2)-\frac{128}{5}\ln(y)\right)y^5$\\
& $+\left(\frac{284664301}{201600}+\frac{28016}{105}\gamma+\frac{14008}{105}\ln(y)+\frac{63472}{105}\ln(2)-\frac{486}{7}\ln(3)-\frac{246367}{1536}\pi^2\right) y^6$\\
&$-\frac{27392}{525}\pi y^{13/2}$\\
&$+\left(\frac{5044}{405}\ln(y)-\frac{413480}{567}\ln(2)+\frac{4617}{7}\ln(3)+\frac{10088}{405}\gamma+\frac{22848244687}{7257600}+\frac{2800873}{131072}\pi^4-\frac{608698367}{884736}\pi^2\right) y^7$\\
&$+\frac{1025002}{3675}\pi y^{15/2}$\\
&$+\left(-\frac{29335719926}{5457375}\gamma-\frac{14667859963}{5457375}\ln(y)-\frac{1836927775597}{1238630400}\pi^2-\frac{830502449}{8388608}\pi^4+\frac{6081934672471237}{195592320000}+\frac{876544}{525}\ln(2)^2\right. $\\
& $-\frac{4096}{5}\zeta(3)+\frac{219136}{525}\gamma^2+\frac{54784}{525}\ln(y)^2-\frac{1953125}{9504}\ln(5)+\frac{876544}{525}\ln(2)\gamma+\frac{438272}{525}\ln(y)\ln(2)$\\
&$\left.+\frac{219136}{525}\gamma\ln(y)-\frac{38722023302}{5457375}\ln(2)-\frac{3572343}{1760}\ln(3)\right)y^8$ \\
&$-\frac{70898413}{3274425}\pi y^{17/2}$\\
&$+\left(\frac{1198510638937}{99324225}\gamma+\frac{1193425238617}{198648450}\ln(y)+\frac{53276112149251}{46242201600}\pi^2+\frac{23033337928985}{3221225472}\pi^4\right. $\\
& $+\frac{37908}{49}\ln(3)^2-\frac{199442049212246428877}{284782417920000} +\frac{37908}{49}\ln(3)\ln(y)+\frac{75816}{49}\ln(3)\ln(2)+\frac{75816}{49}\ln(3)\gamma $\\
& $-\frac{644800}{63}\ln(2)^2+\frac{304256}{105}\zeta(3)-\frac{21788992}{11025}\gamma^2-\frac{5447248}{11025}\ln(y)^2+\frac{2283203125}{741312}\ln(5)-\frac{34759552}{3675}\ln(2)\gamma
$\\
& $\left.-\frac{17379776}{3675}\ln(y)\ln(2)-\frac{21788992}{11025}\gamma\ln(y)+\frac{11647126988311}{496621125}\ln(2)-\frac{325284577623}{35672000}\ln(3)\right)y^9$\\
& $+\left(-\frac{438272}{1575}\pi^2+\frac{46895104}{55125}\gamma-\frac{3008350528127363}{524431908000}+\frac{23447552}{55125}\ln(y)+\frac{93790208}{55125}\ln(2)\right) \pi y^{19/2}
+O(y^{10})$\\
\hline
$h_{kk\,\hat s}^{\rm R}$ & $-y^{5/2}+\frac92 y^{7/2}-\frac38 y^{9/2}+\left(\frac{717}{16}-\frac{87}{32}\pi^2\right)y^{11/2}+(\frac{2560511}{9600}-\frac{5277}{128}\pi^2+\frac{672}{5}\gamma+\frac{336}{5}\ln(y)+\frac{4064}{15}\ln(2)) y^{13/2}$\\
&$+\left(\frac{1564965039}{313600}-\frac{5076}{35}\ln(y)-\frac{2728}{3}\ln(2)+\frac{2187}{7}\ln(3)-\frac{10152}{35}\gamma-\frac{554257}{1024}\pi^2\right) y^{15/2}$\\
&$+\frac{177104}{1575}\pi y^8$\\
&$+\left(-\frac{837392}{405}\gamma-\frac{418696}{405}\ln(y)+\frac{127286591}{221184}\pi^2+\frac{16831801}{983040}\pi^4-\frac{785299919981}{304819200}-\frac{439984}{567}\ln(2)-\frac{181521}{70}\ln(3)\right) y^{17/2}$\\
&$-\frac{634742}{2205}\pi y^9$\\
&$+\left(\frac{4604720159}{363825}\gamma+\frac{4604720159}{727650}\ln(y)+\frac{133780168623}{18350080}\pi^2+\frac{94225974289}{100663296}\pi^4-\frac{434560875210242483}{2581818624000}-\frac{2335168}{525}\ln(2)^2\right.$\\
&$+2176\zeta(3)-\frac{116416}{105}\gamma^2-\frac{29104}{105}\ln(y)^2+\frac{48828125}{28512}\ln(5)-\frac{999808}{225}\ln(2)\gamma-\frac{499904}{225}\ln(y)\ln(2)-\frac{116416}{105}\gamma\ln(y)$\\
&$\left.+\frac{7285095463}{779625}\ln(2)+\frac{72281079}{12320}\ln(3)\right) y^{19/2}
+O(y^{10})$\\
\hline
$h_{kk\,\hat s^2}^{\rm R}$ & $ -\frac12 C_Q y^3+(-\frac74 C_Q-1) y^4+(-\frac72+\frac{213}{16} C_Q) y^5+
\left(\frac{117}{8}+\frac{2641}{1024}\pi^2 C_Q-\frac{6929}{96} C_Q\right) y^6$\\
&$+(\frac{1134821}{6400} C_Q-\frac{1024}{5}\ln(2) C_Q-\frac{5469}{1024}\pi^2 C_Q-\frac{87}{32}\pi^2+\frac{1045}{16}-\frac{256}{5}\ln(y) C_Q-\frac{512}{5}\gamma C_Q) y^7$\\
&$+\left(\frac{768240591}{627200} C_Q-\frac{128}{5}\gamma+\frac{114488}{105}\ln(2) C_Q+\frac{15336}{35}\gamma C_Q+\frac{7668}{35}\ln(y) C_Q-\frac{12115}{256}\pi^2-\frac{896}{15}\ln(2)\right. $\\
&$\left.-\frac{64}{5}\ln(y)+\frac{5120431}{9600}-\frac{1458}{7}\ln(3) C_Q-\frac{9227583}{65536}\pi^2  C_Q\right) y^8$\\
&$-\frac{48064}{525} C_Q\pi y^{17/2}$\\
&$+\left(-448\ln(2)-\frac{7904}{35}\ln(y)-\frac{472804}{405}\ln(2) C_Q+\frac{14094}{7}\ln(3) C_Q+\frac{940066}{2835}\ln(y) C_Q+\frac{253551299}{31457280}\pi^4 C_Q\right. $\\
&$\left.+\frac{8030900799083}{609638400} C_Q-\frac{15808}{35}\gamma-\frac{2916}{7}\ln(3)-\frac{25391054267}{14155776}\pi^2 C_Q+\frac{1880132}{2835}\gamma C_Q+\frac{228719}{1024}\pi^2-\frac{1016803379}{940800}\right) y^9$\\
&$+(\frac{14272}{225}\pi+\frac{1861231}{4410} C_Q\pi) y^{19/2}
+O(y^{10})$\\
\hline
$M[\partial_r h_{kk\,(0)}]^{\rm R}$& $y^2-\frac{13}{2} y^3+\frac{75}{8} y^4+\left(-\frac{585}{16}+\frac{87}{32}\pi^2\right) y^5$\\
&$+\left(\frac{1208369}{9600}-\frac{256}{5}\ln(y)-\frac{1024}{5}\ln(2)-\frac{512}{5}\gamma+\frac{637}{512}\pi^2\right) y^6$\\
&$+\left(\frac{181737541}{313600}+\frac{920}{7}\ln(y)+\frac{1840}{7}\gamma+\frac{5168}{7}\ln(2)-\frac{1458}{7}\ln(3)-\frac{37959}{1024}\pi^2\right) y^7-\frac{41344}{525}\pi y^{15/2}$\\
&$+\left(\frac{3611672}{2835}\gamma+\frac{1805836}{2835}\ln(y)-\frac{1924352635}{1769472}\pi^2-\frac{46207399}{3932160}\pi^4+\frac{1064408}{2835}\ln(2)+1701\ln(3)+\frac{2615606254229}{304819200}\right) y^8$\\
&$+\frac{181991}{735}\pi y^{17/2}$\\
&$+\left(-\frac{52010166562}{5457375}\gamma-\frac{26005083281}{5457375}\ln(y)-\frac{9765625}{9504}\ln(5)+\frac{438272}{525}\gamma\ln(y)+\frac{1753088}{525}\ln(2)\gamma\right. $\\
&$+\frac{876544}{525}\ln(y)\ln(2)-\frac{12791999371003}{2477260800}\pi^2+\frac{3809709473}{251658240}\pi^4-\frac{49655718274}{5457375}\ln(2)-\frac{9225009}{2464}\ln(3)$\\
&$\left.+\frac{717160650537503299}{12909093120000}-\frac{8192}{5}\zeta(3)+\frac{109568}{525}\ln(y)^2+\frac{1753088}{525}\ln(2)^2+\frac{438272}{525}\gamma^2\right) y^9$\\
&$+\frac{6359258497}{6548850}\pi y^{19/2}
+O(y^{10})$\\
\hline
$\tilde\Omega_{1{\hat s}}^{\rm R}$& $-\frac{13}{4} y^{5/2}+\frac{45}{8} y^{7/2}+\frac{209}{32} y^{9/2}+\left(-32\gamma-16\ln(y)-\frac{992}{15}\ln(2)+\frac{9161}{64}-\frac{21625}{2048}\pi^2\right) y^{11/2}$\\
&$+\left(-\frac{34586059}{268800}+\frac{94459}{2048}\pi^2+\frac{1196}{35}\ln(2)-\frac{332}{5}\gamma-\frac{166}{5}\ln(y)-\frac{2187}{14}\ln(3)\right) y^{13/2}-\frac{6032}{1575}\pi y^7$\\
&$+\left(\frac{100568}{105}\gamma+\frac{14648}{105}\ln(2)+\frac{84807}{70}\ln(3)-\frac{101936324827}{6773760}+\frac{371379453}{262144}\pi^2+\frac{1407987}{1048576}\pi^4+\frac{50284}{105}\ln(y)\right)y^{15/2}
+O(y^8)$\\
\end{tabular}
\end{ruledtabular}
\end{table*}

\subsection{Final result for $z_1^{(1)\hat s^2}(y)$}

Individual SF computations of the various terms give
\begin{widetext}
\begin{eqnarray}
h_{kk\,(0)}^{\rm R}(y)&=&
-2y+5y^2+\frac54y^3+\left(-\frac{1261}{24}+\frac{41}{16}\pi^2\right)y^4
+O(y^5)\,,\nonumber\\
M[\partial_r h_{kk\,(0)}]^{\rm R}(y)&=&
y^2-\frac{13}{2}y^3+\frac{75}{8}y^4+\left(-\frac{585}{16}+\frac{87}{32}\pi^2\right)y^5
+O(y^6)\,,\nonumber\\
\tilde\Omega_{1{\hat s}}^{\rm R}(y)&=&
-\frac{13}{4}y^{5/2}+\frac{45}{8}y^{7/2}+\frac{209}{32}y^{9/2}
+O(y^{11/2})
\,,\nonumber\\
h_{kk\,\hat s^2}^{\rm R}(y)&=&
-\frac12C_Qy^3+\left(-\frac{7}{4}C_Q-1\right)y^4+\left(\frac{213}{16}C_Q-\frac72\right)y^5\nonumber\\
&&
+\left[\frac{117}{8}+\left(-\frac{6929}{96}+\frac{2641}{1024}\pi^2\right)C_Q\right]y^6
+O(y^7)\,,
\end{eqnarray}
\end{widetext}
so that 
\begin{eqnarray}
\label{z11SFsquad}
z_1^{(1)\hat s^2}(y)&=&C_Q\left[
\frac12 y^3-\frac12 y^4-\frac54 y^5\right. \nonumber\\
&+&\left. \left(\frac{62}{3}-\frac{1249}{2048}\pi^2\right)y^6\right]
+O(y^7)\,.
\end{eqnarray}

Including the MST solutions we finally obtain
\begin{widetext}
\begin{eqnarray}
\label{eq:z1_squad_final}
z_1^{(1)\hat s^2}(y)&=&C_Q\left[
\frac12 y^3-\frac12 y^4-\frac54 y^5
+\left(\frac{62}{3}-\frac{1249}{2048}\pi^2\right) y^6\right.\nonumber\\
&&
+\left(\frac{2897}{4096}\pi^2-\frac{2573}{300}+\frac{128}{5}\gamma+\frac{64}{5}\ln(y)+\frac{256}{5}\ln(2)\right) y^7\nonumber\\
&&
+\left(-\frac{156143}{300}+\frac{7980487}{131072}\pi^2-\frac{19048}{105}\ln(2)-\frac{2232}{35}\gamma-\frac{1116}{35}\ln(y)+\frac{729}{14}\ln(3)\right) y^8\nonumber\\
&&
+\frac{13696}{525}\pi  y^{17/2}\nonumber\\
&&
+\left(-\frac{311081711}{56700}+\frac{20797457131}{28311552}\pi^2+\frac{266248}{2835}\ln(2)-\frac{599288}{2835}\gamma-\frac{299644}{2835}\ln(y)-\frac{5589}{14}\ln(3)\right.\nonumber\\
&&\left.\left.
-\frac{29225393}{4194304}\pi^4\right) y^9
-\frac{31777}{450}\pi  y^{19/2}\right]
+O(y^{10})\,,
\end{eqnarray}
\end{widetext}
which is proportional to $C_Q$. 
Therefore, the ratio between $z_1^{(1)\hat s^2}$ and its limiting value in the black hole case ($C_Q=1$) is exactly equal to the polarizability parameter, allowing to discriminate the nature of the extended body (either a black hole or a neutron star) and its equation of state.

\section{The PN expectation}

Using a PN Hamiltonian $H$ for the two-body system, the redshift invariant $z_1$ of body 1 to linear order in spin can be calculated from
\beq
z_1= \frac{\partial H(x^i, p_i, S_1^i, S_2^i; m_1, m_2)}{\partial m_1} \,,
\eeq
which follows from the \lq\lq first law" of two-body dynamics \cite{Blanchet:2012at}.
In order to extend this formula to quadratic order in spin, one must add to the Lagrangian in Eq. (3.2) of Ref. \cite{Blanchet:2012at} the spin-induced (SI) quadrupole interactions \cite{Porto:2008jj,Levi:2015ixa},
\beq
  L_\text{SI}^{SS} \sim - \sum_{A=1}^2 \frac{C_{A(ES^2)}}{2 m_A} R_{\mu\alpha\nu\beta} U^\alpha U^\beta S(U)^\mu S(U)^\nu \,,
\eeq
where $A = 1,2$ labels the two bodies of the binary, and $U$ denotes the unit tangent vector to the center of mass world line.
Now, if one takes $\bar C_{A(ES^2)} = C_{A(ES^2)} / m_A$ (instead of just $C_{A(ES^2)}$) as constant when varying the masses $m_A$, then the contribution $L_\text{SI}$ is in fact irrelevant.
That is, the arguments in Sec. III of Ref. \cite{Blanchet:2012at} leading to the formula for the redshift apply unchanged.
The formula for the redshift in the presence of spin-induced (SI) quadrupole interactions at quadratic order in spin hence reads
\beq
z_1= \frac{\partial H(x^i, p_i, S_1^i, S_2^i; m_1, m_2, \bar C_{1(ES^2)}, \bar C_{2(ES^2)})}{\partial m_1} \,.
\eeq
Using known results for the PN dynamics at quadratic order in spin from LO \cite{Barker:1975ae,DEath:1975wqz,Barker:OConnell:1979,Poisson:1997ha,Thorne:1984mz}, NLO \cite{Steinhoff:2007mb,Porto:2006bt,Porto:2008tb,Levi:2008nh,Porto:2008jj,Steinhoff:2008ji,Hergt:2010pa,Hergt:2011ik,Bohe:2015ana}, and NNLO \cite{Hartung:2011ea,Levi:2011eq,Hartung:2013dza,Levi:2014sba,Levi:2015ixa}, as summarized by the Hamiltonian in Eqs. (4.29)-(4.32) of Ref. \cite{Levi:2015uxa} and Eqs. (3.5)-(3.6) of Ref. \cite{Levi:2016ofk}, it is now straightforward to compute the redshift invariant at the second order in spin, $z_1^{\rm SS}$, which is relevant for the present analysis.
Its expression in terms of the frequency-related variable $x=[(m_1+m_2)\Omega]^{2/3}$ (with $\Omega=\frac{\partial H}{\partial p_\phi}$) is the following
\begin{widetext}
\begin{eqnarray}
z_1^{\rm SS}(x;\nu; \chi_1,\chi_2)  &=& \left\{\left(-\frac14  \Delta\nu-\frac12 \nu^2+\frac14  \nu\right)C_{1(ES^2)} \chi_1^2
+\chi_1\chi_2 \nu^2  
+\left[\left(\frac12\nu-\frac14\right)\Delta-\frac14+\nu-\frac12 \nu^2\right] C_{2(ES^2)} \chi_2^2\right\} x^3
\nonumber\\
&+& 
\left\{\left[\left(\frac{17}{18}\nu^2+\frac{4}{9}\nu\right)\Delta-\frac{4}{9}\nu -\frac{14}{9}\nu^2+\frac{7}{9}\nu^3\right.\right.\nonumber\\
&+&\left.
\left(\left(-\frac{7}{24}\nu+\frac{5}{6}\nu^2\right)\Delta-\frac{17}{12}\nu^2+\frac{7}{12}\nu^3 +\frac{7}{24}\nu\right) C_{1(ES^2)}\right]\chi_1^2 \nonumber\\
&+&
\left(\frac{7}{18}\nu^3+\frac{1}{12}\nu^2-\frac{1}{12}\Delta\nu^2  \right)\chi_1\chi_2  \nonumber\\
&+& \left. 
\left[\left(\frac{13}{18}\nu+\frac{1}{3}-\frac{23}{18}\nu^2\right)\Delta +\frac{7}{9}\nu^3-\frac{61}{18}\nu^2+\frac{1}{18}\nu+\frac{1}{3} 
\right.\right.\nonumber\\
&+& \left. \left.
\left(\frac{7}{12}\nu^3 -\frac{7}{8}+\frac{25}{6} \nu-\frac{103}{24}\nu^2+\left(-\frac{7}{8} +\frac{29}{12}\nu-\frac{29}{24}\nu^2\right)\Delta\right)C_{2(ES^2)}\right]\chi_2^2\right\} x^4\nonumber\\
&+& \left\{ \left[\left(\frac{23}{84}\nu-\frac{121}{27}\nu^3+\frac{985}{216}\nu^2\right)\Delta-\frac{23}{84}\nu-\frac{77}{54}\nu^4-\frac{9469}{1512}\nu^2+\frac{9563}{756}\nu^3\right.\right.  \nonumber\\
&+&\left.
\left(-\frac{187}{224}\nu-\frac{11}{144}\nu^4-\frac{299}{504} \nu^2+\frac{12245}{2016}\nu^3+\left(-\frac{325}{288}\nu^3+\frac{2281}{1008}\nu^2+\frac{187}{224} \nu\right)\Delta\right) C_{1(ES^2)}  \right]\chi_1^2\nonumber\\
&+& \left[\left(\frac{53}{72}\nu^3+\frac{143}{48}\nu^2\right)\Delta-\frac{331}{36}\nu^3+\frac{113}{48}\nu^2-\frac{583}{216}\nu^4\right]\chi_1 \chi_2\nonumber\\
&+& 
\left[-\frac{77}{54}\nu^4+\frac{23}{56}+\frac{8887}{378}\nu^3+\frac{1319}{252}\nu-\frac{2963}{108}\nu^2+\left(\frac{142}{27}\nu^3+\frac{23}{56}+\frac{109}{18}\nu-\frac{2741}{189}\nu^2\right)\Delta\right.\nonumber\\
&+&  
\left(-\frac{561}{224}-\frac{11}{144}\nu^4 -\frac{1487}{72}\nu^2+\frac{4799}{504}\nu^3+\frac{4525}{336}\nu\right.\nonumber\\
&+&\left.\left.\left.
\left(-\frac{8815}{1008}\nu^2-\frac{561}{224}+\frac{85}{72}\nu^3+\frac{203}{24} \nu\right)\Delta\right)C_{2(ES^2)} \right]\chi_2^2\right\} x^5 
+O(x^6)\,,
\end{eqnarray}
\end{widetext}
with $\nu=m_1m_2/(m_1+m_2)^2$ and $\Delta \equiv (m_2-m_2)/(m_1+m_2)=\sqrt{1-4\nu}$.
Here we have used the spin-related quantities
\beq
\chi_A = \frac{S_A}{m_A^2}\,,
\eeq
with $\chi_2=0$ in the Schwarzschild case, so that
\beq
\chi_1= \frac{S_1}{m_1^2}=\frac{m_2}{m_1}\frac{S_1}{m_1 m_2}=\frac{1}{q} \hat s\,,
\eeq
with $q=m_1/m_2$, and $C_{1(ES^2)}=1=C_{2(ES^2)}$ in the black hole case (see also Ref. \cite{Hergt:2012zx}).

The corresponding 1SF expansion then reads
\begin{eqnarray}
z_1^{\rm 1SF}(y)&=& 
y-y^2-y^3+\left(\frac{76}{3}-\frac{41}{32}\pi^2\right)y^4+O(y^5)\nonumber\\
&+&
\left(-\frac{7}{3}y^{5/2}-\frac{13}{3}y^{7/2}-23y^{9/2}+O(y^{11/2})\right)\chi_2\nonumber\\
&+&
\left[C_{2(ES^2)}y^3+\left(\frac{17}{9}+\frac{11}{3}C_{2(ES^2)}\right)y^4\right.\nonumber\\
&+&\left. \left(\frac{832}{63}+\frac{215}{21}C_{2(ES^2)}\right)y^5+O(y^6)\right]\chi_2^2
\,,
\end{eqnarray}
where we have introduced the variable $y$ such that $x=y(1+q)^{2/3}$ with $\nu=\frac{q}{(q+1)^2}$, reproducing known results in the black hole case $C_{2(ES^2)}=1$ (see, e.g., Ref. \cite{Bini:2019lcd}) .

In the test-body limit we get
\begin{eqnarray}
z_1^{\rm 0SF}(y)&=&1-\frac32y-\frac98y^2-\frac{27}{16}y^3-\frac{405}{128}y^4\nonumber\\
&&
\left(2y^{5/2}+3y^{7/2}+\frac{27}{4}y^{9/2}\right)\chi_2\nonumber\\
&&
+\left[-\frac12C_{2(ES^2)}y^3 +\left(\frac{2}{3}-\frac{7}{4}C_{2(ES^2)}\right)y^4\right. \nonumber\\
&& \left. +\left(\frac{23}{28}-\frac{561}{112}C_{2(ES^2)}\right)y^5\right]\chi_2^2\,,
\end{eqnarray}
in agreement with the (exact) Kerr result ($\chi_2=\hat a$, $C_{2(ES^2)}=1$)
\beq
z_1^{\rm Kerr}= \frac{(1-3y'+2\hat a y'^{3/2})^{1/2}}{1+\hat a y'^{3/2}}\Bigg|_{y'=y/(1-\hat a y^{3/2})^{2/3}}  \,,
\eeq
namely
\begin{eqnarray}
z_1^{\rm Kerr}
&\approx & 1-\frac32 y+\ldots \nonumber\\
&&+
\left(2 y^{5/2}+3y^{7/2}+\frac{27}{4}y^{9/2}+\ldots \right)\hat a\nonumber\\
&& +\left(-\frac12  y^3 -\frac{13}{12} y^4-\frac{67}{16} y^5\right)\hat a^2 \nonumber\\
&&+O(y^{11/2},\hat a^3)\,,
\end{eqnarray}
in its expanded form.

Let us discuss the same results in terms of the spin variable $\hat s$ instead of $\chi_1$.
In the test-body limit we get
\begin{eqnarray}
z_1^{\rm 0SF}(y)&=&1-\frac32y-\frac98y^2-\frac{27}{16}y^3-\frac{405}{128}y^4\nonumber\\
&&
\left(2y^{5/2}+3y^{7/2}+\frac{27}{4}y^{9/2}\right)\chi_2\nonumber\\
&&
+\left[-\frac12C_{2(ES^2)}y^3 +\left(\frac{2}{3}-\frac{7}{4}C_{2(ES^2)}\right)y^4\right. \nonumber\\
&& \left. +\left(\frac{23}{28}-\frac{561}{112}C_{2(ES^2)}\right)y^5\right]\chi_2^2\nonumber\\
&&
+\left(-\frac{3}{2}y^4-\frac{9}{4}y^5\right)\hat s^2\,,
\end{eqnarray}
which agrees with Eq. \eqref{z10} for $\chi_2=0$ and $C_{1(ES^2)}=C_Q$.

The 1SF expansion is 
\begin{eqnarray}
z_1^{\rm 1SF}(y)&=& 
y-y^2-y^3+\left(\frac{76}{3}-\frac{41}{32}\pi^2\right)y^4+O(y^5)\nonumber\\
&&
+\left(-\frac{7}{3}y^{5/2}-\frac{13}{3}y^{7/2}-23y^{9/2}+O(y^{11/2})\right)\chi_2\nonumber\\
&&
+\left(y^{7/2}-3y^{9/2}\right)\hat s\nonumber\\
&&
+\left[C_{2(ES^2)}y^3+\left(\frac{17}{9}+\frac{11}{3}C_{2(ES^2)}\right)y^4\right. \nonumber\\
&&\left. +\left(\frac{832}{63}+\frac{215}{21}C_{2(ES^2)}\right)y^5+O(y^6)\right]\chi_2^2\nonumber\\
&&
+\left(y^3+\frac{16}{3}y^5\right)\hat s\chi_2\nonumber\\
&&
+\left(\frac12y^3-\frac12y^4-\frac54y^5\right)C_{1(ES^2)}\hat s^2
\,,
\end{eqnarray}
which agrees with Eq. \eqref{z11SFsquad} for $\chi_2=0$ and $C_{1(ES^2)}=C_Q$.

\section{Concluding remarks}

We have studied the perturbations induced by a classical extended object endowed with both dipolar and (spin-induced) quadrupolar structure moving along an equatorial circular orbit on the Schwarzschild background, the spin vector being orthogonal to the motion plane.
The metric perturbations have been obtained by using the standard Teukolsky formalism in a radiation gauge within the framework of first-order gravitational self-force.
We have computed for the first time the spin-squared contribution to Detweiler's redshift invariant at a high-PN order, checking also the agreement of the first terms of the expansion with the corresponding PN expectation.
For the purpose of the latter, we utilize that Detweiler's redshift invariant has a counterpart in the PN Hamiltonian formalism: the PN redshift follows from the \lq\lq first law" of two-body dynamics, which we extended from the linear-in-spin level \cite{Blanchet:2012at} to quadratic level (including spin-induced quadrupole interactions).

The transcription of this new result into other formalisms like the EOB one requires some care, since there is the choice, dictated by the Kerr solution,  to include them (eventually in a resummed form) in the orbital sector of the Hamiltonian (as in Ref. \cite{Balmelli:2015zsa,Damour:2001tu}) and/or in an external spin-squared Hamiltonian (as in Ref. \cite{Barausse:2009xi}).
Following the method in Refs. \cite{Bini:2019nra,Damour:2019lcq,Antonelli:2020aeb} which utilize self-force results to derive new PN results (making crucial use of the mass-ratio dependence of the scattering angle), it is also conceivable that an extension of the results in the present paper to eccentric orbits could be suffucient to derive the NNNLO spin-squared conservative PN Hamiltonian at 5PN for aligned spins (see Ref. \cite{Levi:2020uwu} for partial results), complementing efforts to complete the knowledge of the 5PN order in the nonspinning sector \cite{Foffa:2019hrb,Blumlein:2019zku,Bini:2019nra}.
These problems will be discussed elsewhere.

\section*{Acknowledgments}
DB thanks T. Damour for many useful discussions.

\appendix

\section{Completion of the metric: the non-radiative modes}

The completion of the metric with the addition of the gauge modes is solved here by studying the perturbation equations corresponding to the lowest multipoles $l=0,1$ in a spherical harmonic decomposition of the metric, following the original approach of Zerilli.
The derivation of these equations and the associated solutions (listed below) closely follow what has been done recently case of a spinning particle,  Ref.  \cite{Bini:2018zde}.
We distinguish the case of the monopole, $l=0$ and the dipole $l=1$ (with both its even and odd parts), corresponding to the addition of mass and angular momentum to the background spacetime.

\subsection{The monopole mode $l=0$}

The nonvanishing metric components are 
\beq
qh_{tt}=\frac{fH_0}{\sqrt{4\pi}}\,,\qquad
qh_{rr}=\frac{H_2}{\sqrt{4\pi}f}\,,
\eeq
where the perturbation functions $H_0$ and $H_2$ satisfy the following equations
\begin{eqnarray}
\frac{dH_2}{dr}+\frac{H_2}{rf}&=& A_0 \delta(r-r_0)+A_1 \delta'(r-r_0)\nonumber\\
&&+A_2\delta''(r-r_0)\,,\nonumber\\
\frac{dH_0}{dr}+\frac{H_2}{rf}&=&B_0 \delta(r-r_0)\,,
\end{eqnarray}
with spin-dependent coefficients $A_0$, $A_1$, $A_2$ and $B_0$ listed in Table \ref{A0B0_etc}.
The solution for the monopole perturbation is thus found to be
\begin{widetext}
\begin{eqnarray}
qh_{tt}&=& \frac{2\delta M}{r} \left[\frac{r f}{r_0 f_0} \left(1-\frac{2 r_0-3 M}{ r_0 f_0} M\zeta_K\hat s (1+{\mathcal B}_0\hat s)\right)H(r_0-r)+H(r-r_0)\right]+{\mathcal B}_1\delta M\hat s^2 \delta(r-r_0)\,,\nonumber\\
qh_{rr}&=&   2\frac{\delta M}{ r f^2} H(r-r_0)-\frac{2}{f_0^2}\delta M M \zeta_K \hat s (1+\hat s {\mathcal C}_1) \delta(r-r_0)+{\mathcal C}_2 \delta M\hat s^2\delta'(r-r_0)\,,
\end{eqnarray}
\end{widetext}
to second order in spin, with coefficients
\begin{eqnarray}
\label{B0def}
{\mathcal B}_0&=&
-u_0^{1/2}\frac{C_Q(1-2u_0)^2+u_0(2-3u_0)(3-4u_0)}{2(1-2 u_0)(2-3 u_0)}
\,,\nonumber\\
{\mathcal B}_1&=&
 -\frac13  C_Q u_0^2
\,,
\end{eqnarray}
and
\begin{eqnarray}
{\mathcal C}_1&=&
-u_0^{1/2}\frac{2C_Q(1-u_0)(1-2u_0)+3u_0(3-4u_0)}{6(1-2 u_0) } 
\,,\nonumber\\
{\mathcal C}_2&=&
\frac13C_Q\frac{u_0}{1-2 u_0}
\,.
\end{eqnarray}
Here $H(x)$ denotes the Heaviside step function, $M\zeta_K=u_0^{3/2}$ and the additional contribution $\delta M$ to the mass of the system is given by the conserved Killing energy \eqref{EJspin} of the extended body.

\subsection{The dipole mode $l=1$ (odd)}

The only nonvanishing metric component is 
\beq
qh_{t\phi}=-\sqrt{\frac{3}{4\pi}}h_0^{\rm(odd)}\sin^2\theta
\,,
\eeq
where the perturbation function $h_0^{\rm(odd)}$ satisfies the equation
\begin{eqnarray}
\frac{dh_0}{dr^2}-\frac{2}{r^2}h_0 &=& C_0 \delta(r-r_0)+C_1 \delta'(r-r_0)\nonumber\\
&&+C_2\delta''(r-r_0)\,,
\end{eqnarray}
where the coefficients $C_0$, $C_1$ and $C_2$  are listed in Table \ref{A0B0_etc}.
The solution for the odd dipole perturbation is thus found to be
\begin{widetext}
\beq
qh_{t\phi} = \left\{-2\frac{\delta J}{r}\left[\frac{r^3}{r_0^3} \left(1-\frac32  (r_0-M)\zeta_K\hat s (1+{\mathcal D}_0\hat s)\right) H(r_0-r)+H(r-r_0)\right]+{\mathcal D}_1\delta J\hat s^2\delta(r-r_0)\right\}\sin^2\theta\,,
\eeq
\end{widetext} 
where 
\begin{eqnarray}
\label{D0def}
{\mathcal D}_0 &=& u_0^{1/2}\frac{-2u_0(1-2u_0)C_Q+9u_0-13u_0^2-2}{2(1-u_0)} \nonumber\\
{\mathcal D}_1 &=&  u_0^2(1-2 u_0)C_Q\,,
\end{eqnarray}
and $\delta J$ is given by the conserved Killing angular momentum \eqref{EJspin} of the extended body.

\subsection{Nonradiative part of $h_{kk}$}

The unsubtracted contribution to $h_{kk\,(+)}^{\rm(nonrad)}$ at the location of the extended body due to nonradiative multipoles is then given by
\begin{eqnarray}
h_{kk\,(+)}^{\rm(nonrad)}&=&h_{tt\,(+)}^{\ell=0,1}+2\zeta h_{t\phi\,(+)}^{\ell=0,1}\nonumber\\
&=&\frac{2y(1-4y)}{\sqrt{1-3y}}-2y^{5/2}\sqrt{1-3y}{\hat s}\nonumber\\
&+&
\frac{y^3}{ (1-3 y)^{3/2}}[(1-3y)(8y^2+y-1)\nonumber\\
&+&
(C_Q-1)(24 y^3-18y^2+6y-1)]\hat s^2
\,,\nonumber\\
\end{eqnarray}
to second order in $\hat s$, where we have used the unperturbed relation \eqref{u0vsy0} to replace $u_0$ with $y$.
The contribution $h_{kk\,(-)}^{\rm(nonrad)}$ from the interior metric perturbation is instead given by
\begin{eqnarray}
h_{kk\,(-)}^{\rm(nonrad)}&=&h_{tt\,(-)}^{\ell=0,1}+2\zeta h_{t\phi\,(-)}^{\ell=0,1}\nonumber\\
&=&\frac{2y(1-4y)}{\sqrt{1-3y}}+\frac{6y^{7/2}}{\sqrt{1-3y}}{\hat s}\nonumber\\
&-&
\frac{y^4}{ (1-3 y)^{3/2}}[(1-3y)(5-2y)\nonumber\\
&+&
(C_Q-1)(12y^2-18y+5)]\hat s^2
\,.\nonumber\\
\end{eqnarray}
The final result for the needed left-right average is then
\beq
\langle h_{kk}^{\rm(nonrad)} \rangle = \frac12 \left( h_{kk\,(+)}^{\rm(nonrad)} +  h_{kk\,(-)}^{\rm(nonrad)}\right)\,,
\eeq
with value
\begin{eqnarray}
\label{hkkleq01}
\langle h_{kk}^{\rm(nonrad)} \rangle &=&\frac{2y(1-4y)}{\sqrt{1-3y}}\nonumber\\
&& -\frac{y^{5/2}(1-6y)}{\sqrt{1-3y}}\hat s
\nonumber\\
&& 
+\frac{y^3}{2(1-3 y)^{3/2}}[(1-3y)((10y^2-4y-1))\nonumber\\
&& 
+(C_Q-1)((12y^3+y-1))]\hat s^2
\,.
\end{eqnarray}

% table 3

\begin{table*}
\caption{\label{A0B0_etc} List of the coefficients entering the source terms of the perturbation equations for the low multipoles.}
\begin{ruledtabular}
\begin{tabular}{ll}
Coefficient & expression \\
\hline
$A_0$ & $4\sqrt{\pi}m_0\frac{u_0}{M(1-3u_0)^{1/2}}
\left[1-u_0^{3/2}\frac{2-5u_0}{2(1-2u_0)(1-3u_0)}{\hat s}
+u_0^2\frac{-2(1-3u_0)(36u_0^3-36u_0^2+23u_0-6)C_Q+9u_0(1-2u_0)(15u_0^2-14u_0+4)}{24(1-2u_0)(1-3u_0)^2}{\hat s}^2\right]$ \\ 
$A_1$ & $-4\sqrt{\pi}m_0\frac{u_0^{3/2}}{(1-3u_0)^{1/2}}{\hat s}
\left[1-u_0^{1/2}\frac{(1-3u_0)(3-4u_0)C_Q+9u_0(1-2u_0)}{6(1-3u_0)}{\hat s}\right]$\\
$A_2$ &$ \frac23\sqrt{\pi}m_0MC_Qu_0\frac{1-2u_0}{(1-3u_0)^{1/2}}{\hat s}^2$\\
$B_0$ & $4\sqrt{\pi}m_0\frac{u_0^{5/2}(1-3u_0)^{1/2}}{M(1-2u_0)}{\hat s}
\left[1+u_0^{3/2}\frac{2C_Q-9(1-2u_0)}{6(1-3u_0)}{\hat s}\right]$\\
\hline
\hline
$C_0$ &$-4\sqrt{3\pi}m_0\frac{u_0^{3/2}}{M(1-3u_0)^{1/2}}
\left[1-u_0^{3/2}\frac{5-12u_0}{2(1-3u_0)}{\hat s}
+u_0^2\frac{2(1-3u_0)(12u_0^2-46u_0+17)C_Q+9u_0(36u_0^2-30u_0+7)}{24(1-3u_0)^2}{\hat s}^2\right]$\\
$C_1$ &$2\sqrt{3\pi}m_0\frac{u_0(1-u_0)}{(1-3u_0)^{1/2}}{\hat s}
\left[1-u_0^{3/2}\frac{2(1-2u_0)(1-3u_0)C_Q+3u_0(3-7u_0)}{2(1-u_0)(1-3u_0)}{\hat s}\right]$\\
$C_2$ &$-\frac23\sqrt{3\pi}m_0MC_Qu_0^{3/2}\frac{1-2u_0}{(1-3u_0)^{1/2}}{\hat s}^2$\\ 
\hline
\end{tabular}
\end{ruledtabular}
\end{table*}


\begin{thebibliography}{99}


%\cite{TheLIGOScientific:2017qsa}
\bibitem{TheLIGOScientific:2017qsa} 
  B.~P.~Abbott {\it et al.} [LIGO Scientific and Virgo Collaborations],
  ``GW170817: Observation of Gravitational Waves from a Binary Neutron Star Inspiral,''
  Phys.\ Rev.\ Lett.\  {\bf 119}, no. 16, 161101 (2017)
  doi:10.1103/PhysRevLett.119.161101
  [arXiv:1710.05832 [gr-qc]].

%\cite{Abbott:2020uma}
\bibitem{Abbott:2020uma} 
  B.~P.~Abbott {\it et al.} [LIGO Scientific and Virgo Collaborations],
  ``GW190425: Observation of a Compact Binary Coalescence with Total Mass $\sim 3.4 M_{\odot}$,''
  arXiv:2001.01761 [astro-ph.HE].
	
%\cite{Abbott:2018wiz}
\bibitem{Abbott:2018wiz} 
  B.~P.~Abbott {\it et al.} [LIGO Scientific and Virgo Collaborations],
  ``Properties of the binary neutron star merger GW170817,''
  Phys.\ Rev.\ X {\bf 9}, no. 1, 011001 (2019)
  doi:10.1103/PhysRevX.9.011001
  [arXiv:1805.11579 [gr-qc]].
		
%\cite{Abbott:2018exr}
\bibitem{Abbott:2018exr} 
  B.~P.~Abbott {\it et al.} [LIGO Scientific and Virgo Collaborations],
  ``GW170817: Measurements of neutron star radii and equation of state,''
  Phys.\ Rev.\ Lett.\  {\bf 121}, no. 16, 161101 (2018)
  doi:10.1103/PhysRevLett.121.161101
  [arXiv:1805.11581 [gr-qc]].
		
%\cite{Poisson:1997ha}
\bibitem{Poisson:1997ha} 
  E.~Poisson,
  ``Gravitational waves from inspiraling compact binaries: The Quadrupole moment term,''
  Phys.\ Rev.\ D {\bf 57}, 5287 (1998)
  doi:10.1103/PhysRevD.57.5287
  [gr-qc/9709032].
		
%\cite{Harry:2018hke}
\bibitem{Harry:2018hke} 
  I.~Harry and T.~Hinderer,
  ``Observing and measuring the neutron-star equation-of-state in spinning binary neutron star systems,''
  Class.\ Quant.\ Grav.\  {\bf 35}, no. 14, 145010 (2018)
  doi:10.1088/1361-6382/aac7e3
  [arXiv:1801.09972 [gr-qc]].
		
%\cite{Laarakkers:1997hb}
\bibitem{Laarakkers:1997hb} 
  W.~G.~Laarakkers and E.~Poisson,
  ``Quadrupole moments of rotating neutron stars,''
  Astrophys.\ J.\  {\bf 512}, 282 (1999)
  doi:10.1086/306732
  [gr-qc/9709033].
	
 %\cite{Krishnendu:2017shb}
\bibitem{Krishnendu:2017shb} 
  N.~V.~Krishnendu, K.~G.~Arun and C.~K.~Mishra,
  ``Testing the binary black hole nature of a compact binary coalescence,''
  Phys.\ Rev.\ Lett.\  {\bf 119}, no. 9, 091101 (2017)
  doi:10.1103/PhysRevLett.119.091101
  [arXiv:1701.06318 [gr-qc]].

%\cite{Krishnendu:2019tjp}
\bibitem{Krishnendu:2019tjp} 
  N.~V.~Krishnendu, M.~Saleem, A.~Samajdar, K.~G.~Arun, W.~Del Pozzo and C.~K.~Mishra,
  ``Constraints on the binary black hole nature of GW151226 and GW170608 from the measurement of spin-induced quadrupole moments,''
  Phys.\ Rev.\ D {\bf 100}, no. 10, 104019 (2019)
  doi:10.1103/PhysRevD.100.104019
  [arXiv:1908.02247 [gr-qc]].
		
\bibitem{Mathisson:1937zz} 
  M.~Mathisson,
  ``Neue mechanik materieller systemes,''
  Acta Phys.\ Polon.\  {\bf 6}, 163 (1937).
	
\bibitem{Papapetrou:1951pa} 
  A.~Papapetrou,
  ``Spinning test particles in general relativity. 1.,''
  Proc.\ Roy.\ Soc.\ Lond.\ A {\bf 209}, 248 (1951).
  doi:10.1098/rspa.1951.0200
	
\bibitem{Dixon:1970zza} 
  W.~G.~Dixon,
  ``Dynamics of extended bodies in general relativity. I. Momentum and angular momentum,''
  Proc.\ Roy.\ Soc.\ Lond.\ A {\bf 314}, 499 (1970).
  doi:10.1098/rspa.1970.0020

%\cite{Bini:2008zzc}
\bibitem{Bini:2008zzc} 
  D.~Bini, P.~Fortini, A.~Geralico and A.~Ortolan,
  ``Quadrupole effects on the motion of extended bodies in Schwarzschild spacetime,''
  Class.\ Quant.\ Grav.\  {\bf 25}, 035005 (2008)
  doi:10.1088/0264-9381/25/3/035005
  [arXiv:0910.2841 [gr-qc]].

%\cite{Bini:2008zzf}
\bibitem{Bini:2008zzf} 
  D.~Bini, P.~Fortini, A.~Geralico and A.~Ortolan,
  ``Quadrupole effects on the motion of extended bodies in Kerr spacetime,''
  Class.\ Quant.\ Grav.\  {\bf 25}, 125007 (2008)
  doi:10.1088/0264-9381/25/12/125007
  [arXiv:0910.2842 [gr-qc]].

\bibitem{Steinhoff:2009tk} 
  J.~Steinhoff and D.~Puetzfeld,
   ``Multipolar equations of motion for extended test bodies in General Relativity,''
  Phys.\ Rev.\ D {\bf 81}, 044019 (2010)
  doi:10.1103/PhysRevD.81.044019
  [arXiv:0909.3756 [gr-qc]].

%\cite{Steinhoff:2012rw}
\bibitem{Steinhoff:2012rw} 
  J.~Steinhoff and D.~Puetzfeld,
  ``Influence of internal structure on the motion of test bodies in extreme mass ratio situations,''
  Phys.\ Rev.\ D {\bf 86}, 044033 (2012)
  doi:10.1103/PhysRevD.86.044033
  [arXiv:1205.3926 [gr-qc]].
	
%\cite{Bini:2013nw}
\bibitem{Bini:2013nw} 
  D.~Bini and A.~Geralico,
  ``Dynamics of quadrupolar bodies in a Schwarzschild spacetime,''
  Phys.\ Rev.\ D {\bf 87}, no. 2, 024028 (2013)
  doi:10.1103/PhysRevD.87.024028
  [arXiv:1408.5261 [gr-qc]].
	
%\cite{Bini:2013uwa}
\bibitem{Bini:2013uwa} 
  D.~Bini and A.~Geralico,
  ``Deviation of quadrupolar bodies from geodesic motion in a Kerr spacetime,''
  Phys.\ Rev.\ D {\bf 89}, no. 4, 044013 (2014)
  doi:10.1103/PhysRevD.89.044013
  [arXiv:1311.7512 [gr-qc]].
	
%\cite{Bini:2014xyr}
\bibitem{Bini:2014xyr} 
  D.~Bini and A.~Geralico,
  ``Extended bodies in a Kerr spacetime: exploring the role of a general quadrupole tensor,''
  Class.\ Quant.\ Grav.\  {\bf 31}, 075024 (2014)
  doi:10.1088/0264-9381/31/7/075024
  [arXiv:1408.5484 [gr-qc]].
		
%\cite{Bini:2014epa}
\bibitem{Bini:2014epa} 
  D.~Bini and A.~Geralico,
  ``Effect of an arbitrary spin orientation on the quadrupolar structure of an extended body in a Schwarzschild spacetime,''
  Phys.\ Rev.\ D {\bf 91}, 104036 (2015)
  doi:10.1103/PhysRevD.91.104036
  [arXiv:1412.7643 [gr-qc]].

\bibitem{Bini:2015zya} 
  D.~Bini, G.~Faye and A.~Geralico,
  ``Dynamics of extended bodies in a Kerr spacetime with spin-induced quadrupole tensor,''
  Phys.\ Rev.\ D {\bf 92}, no. 10, 104003 (2015)
  doi:10.1103/PhysRevD.92.104003
  [arXiv:1507.07441 [gr-qc]].

%\cite{Barausse:2009aa}
\bibitem{Barausse:2009aa} 
  E.~Barausse, E.~Racine and A.~Buonanno,
  ``Hamiltonian of a spinning test-particle in curved spacetime,''
  Phys.\ Rev.\ D {\bf 80}, 104025 (2009)
  Erratum: [Phys.\ Rev.\ D {\bf 85}, 069904 (2012)]
  doi:10.1103/PhysRevD.85.069904, 10.1103/PhysRevD.80.104025
  [arXiv:0907.4745 [gr-qc]].

\bibitem{Kibble:1963} 
  T.~W.~B.~Kibble,
  ``Canonical Variables for interacting gravitational and Dirac fields,''
  J.\ Math.\ Phys. {\bf 4}, 1433 (1963)
  doi:10.1063/1.1703923.

%\cite{Khriplovich:1989kg}
\bibitem{Khriplovich:1989kg} 
  I.~B.~Khriplovich,
  ``Spinning Particle In A Gravitational Field,''
  Sov.\ Phys.\ JETP. {\bf 69}, 217 (1989).

%\cite{Witzany:2018ahb}
\bibitem{Witzany:2018ahb} 
  V.~Witzany, J.~Steinhoff and G.~Lukes-Gerakopoulos,
  ``Hamiltonians and canonical coordinates for spinning particles in curved space-time,''
  Class.\ Quant.\ Grav.\  {\bf 36}, no. 7, 075003 (2019)
  doi:10.1088/1361-6382/ab002f
  [arXiv:1808.06582 [gr-qc]].

%\cite{Vines:2016unv}
\bibitem{Vines:2016unv} 
  J.~Vines, D.~Kunst, J.~Steinhoff and T.~Hinderer,
  ``Canonical Hamiltonian for an extended test body in curved spacetime: To quadratic order in spin,''
  Phys.\ Rev.\ D {\bf 93}, no. 10, 103008 (2016)
  doi:10.1103/PhysRevD.93.103008
  [arXiv:1601.07529 [gr-qc]].

%\cite{Blanchet:2013haa}
\bibitem{Blanchet:2013haa}
  L.~Blanchet,
  ``Gravitational Radiation from Post-Newtonian Sources and Inspiralling Compact Binaries,''
  Living Rev.\ Rel.\  {\bf 17}, 2 (2014)
  doi:10.12942/lrr-2014-2
  [arXiv:1310.1528 [gr-qc]].

%\cite{Schafer:2018kuf}
\bibitem{Schafer:2018kuf} 
  G.~Schäfer and P.~Jaranowski,
  ``Hamiltonian formulation of general relativity and post-Newtonian dynamics of compact binaries,''
  Living Rev.\ Rel.\  {\bf 21}, no. 1, 7 (2018)
  doi:10.1007/s41114-018-0016-5
  [arXiv:1805.07240 [gr-qc]].

%\cite{Pati:2000vt}
\bibitem{Pati:2000vt}
  M.~E.~Pati and C.~M.~Will,
  ``Post-Newtonian gravitational radiation and equations of motion via direct integration of the relaxed Einstein equations. 1. Foundations,''
  Phys.\ Rev.\ D {\bf 62}, 124015 (2000)
  doi:10.1103/PhysRevD.62.124015
  [gr-qc/0007087].

%\cite{Futamase:2007zz}
\bibitem{Futamase:2007zz}
  T.~Futamase and Y.~Itoh,
  ``The post-Newtonian approximation for relativistic compact binaries,''
  Living Rev.\ Rel.\  {\bf 10}, 2 (2007).
  doi:10.12942/lrr-2007-2.

%\cite{Bel:1981be}
\bibitem{Bel:1981be} 
  L.~Bel, T.~Damour, N.~Deruelle, J.~Ibanez and J.~Martin,
   ``Poincaré-invariant gravitational field and equations of motion of two pointlike objects: The postlinear approximation of general relativity,''
  Gen.\ Rel.\ Grav.\  {\bf 13}, 963 (1981).
  doi:10.1007/BF00756073

\bibitem{Westpfahl:1985}
  K.~Westpfahl,
  ``High-Speed Scattering of Charged andUncharged Particles in General Relativity,''
  Fortschritte der Physik {\bf 33} 417 (1985)
  doi:10.1002/prop.2190330802.

%\cite{Goldberger:2007hy}
\bibitem{Goldberger:2007hy} 
  W.~D.~Goldberger,
  ``Les Houches lectures on effective field theories and gravitational radiation,''
  hep-ph/0701129.

%\cite{Rothstein:2014sra}
\bibitem{Rothstein:2014sra} 
  I.~Z.~Rothstein,
  ``Progress in effective field theory approach to the binary inspiral problem,''
  Gen.\ Rel.\ Grav.\  {\bf 46}, 1726 (2014).
  doi:10.1007/s10714-014-1726-y

%\cite{Cheung:2018wkq}
\bibitem{Cheung:2018wkq} 
  C.~Cheung, I.~Z.~Rothstein and M.~P.~Solon,
  ``From Scattering Amplitudes to Classical Potentials in the Post-Minkowskian Expansion,''
  Phys.\ Rev.\ Lett.\  {\bf 121}, no. 25, 251101 (2018)
  doi:10.1103/PhysRevLett.121.251101
  [arXiv:1808.02489 [hep-th]].

\bibitem{Detweiler:2008ft} 
  S.~L.~Detweiler,
  ``A consequence of the gravitational self-force for circular orbits of the Schwarzschild geometry,''
  Phys.\ Rev.\ D {\bf 77}, 124026 (2008)
  doi:10.1103/PhysRevD.77.124026
  [arXiv:0804.3529 [gr-qc]].

%\cite{Barack:2009ux}
\bibitem{Barack:2009ux} 
  L.~Barack,
   ``Gravitational self force in extreme mass-ratio inspirals,''
  Class.\ Quant.\ Grav.\  {\bf 26}, 213001 (2009)
  doi:10.1088/0264-9381/26/21/213001
  [arXiv:0908.1664 [gr-qc]].

\bibitem{Bini:2013zaa} 
  D.~Bini and T.~Damour,
   ``Analytical determination of the two-body gravitational interaction potential at the fourth post-Newtonian approximation,''
  Phys.\ Rev.\ D {\bf 87}, no. 12, 121501 (2013)
  doi:10.1103/PhysRevD.87.121501
  [arXiv:1305.4884 [gr-qc]].

%\cite{Buonanno:1998gg}
\bibitem{Buonanno:1998gg} 
  A.~Buonanno and T.~Damour,
   ``Effective one-body approach to general relativistic two-body dynamics,''
  Phys.\ Rev.\ D {\bf 59}, 084006 (1999)
  doi:10.1103/PhysRevD.59.084006
  [gr-qc/9811091].

%\cite{Buonanno:2000ef}
\bibitem{Buonanno:2000ef} 
  A.~Buonanno and T.~Damour,
   ``Transition from inspiral to plunge in binary black hole coalescences,''
  Phys.\ Rev.\ D {\bf 62}, 064015 (2000)
  doi:10.1103/PhysRevD.62.064015
  [gr-qc/0001013].

%\cite{Damour:2014jta}
\bibitem{Damour:2014jta} 
  T.~Damour, P.~Jaranowski and G.~Schäfer,
   ``Nonlocal-in-time action for the fourth post-Newtonian conservative dynamics of two-body systems,''
  Phys.\ Rev.\ D {\bf 89}, no. 6, 064058 (2014)
  doi:10.1103/PhysRevD.89.064058
  [arXiv:1401.4548 [gr-qc]].
 
%\cite{Damour:2015isa}
\bibitem{Damour:2015isa} 
  T.~Damour, P.~Jaranowski and G.~Schäfer,
   ``Fourth post-Newtonian effective one-body dynamics,''
  Phys.\ Rev.\ D {\bf 91}, no. 8, 084024 (2015)
  doi:10.1103/PhysRevD.91.084024
  [arXiv:1502.07245 [gr-qc]].

%\cite{Bernard:2016wrg}
\bibitem{Bernard:2016wrg} 
  L.~Bernard, L.~Blanchet, A.~Bohé, G.~Faye and S.~Marsat,
  ``Energy and periastron advance of compact binaries on circular orbits at the fourth post-Newtonian order,''
  Phys.\ Rev.\ D {\bf 95}, no. 4, 044026 (2017)
  doi:10.1103/PhysRevD.95.044026
  [arXiv:1610.07934 [gr-qc]].
 
%\cite{Bernard:2017ktp}
\bibitem{Bernard:2017ktp} 
  L.~Bernard, L.~Blanchet, G.~Faye and T.~Marchand,
   ``Center-of-Mass Equations of Motion and Conserved Integrals of Compact Binary Systems at the Fourth Post-Newtonian Order,''
  Phys.\ Rev.\ D {\bf 97}, no. 4, 044037 (2018)
  doi:10.1103/PhysRevD.97.044037
  [arXiv:1711.00283 [gr-qc]].

%\cite{Foffa:2019yfl}
\bibitem{Foffa:2019yfl} 
  S.~Foffa, R.~A.~Porto, I.~Rothstein and R.~Sturani,
  ``Conservative dynamics of binary systems to fourth Post-Newtonian order in the EFT approach II: Renormalized Lagrangian,''
  Phys.\ Rev.\ D {\bf 100}, no. 2, 024048 (2019)
  doi:10.1103/PhysRevD.100.024048
  [arXiv:1903.05118 [gr-qc]].

%\cite{Blumlein:2020pog}
\bibitem{Blumlein:2020pog} 
  J.~Blümlein, A.~Maier, P.~Marquard and G.~Schäfer,
  ``Fourth post-Newtonian Hamiltonian dynamics of two-body systems from an effective field theory approach,''
  arXiv:2003.01692 [gr-qc].

%\cite{Antonelli:2020aeb}
\bibitem{Antonelli:2020aeb} 
  A.~Antonelli, C.~Kavanagh, M.~Khalil, J.~Steinhoff and J.~Vines,
  ``Gravitational spin-orbit coupling through third-subleading post-Newtonian order: from first-order self-force to arbitrary mass ratios,''
  arXiv:2003.11391 [gr-qc].

%\cite{Levi:2020kvb}
\bibitem{Levi:2020kvb} 
  M.~Levi, A.~J.~Mcleod and M.~Von Hippel,
  ``N$^3$LO gravitational spin-orbit coupling at order $G^4$,''
  arXiv:2003.02827 [hep-th].

%\cite{Hartung:2011ea}
\bibitem{Hartung:2011ea} 
  J.~Hartung and J.~Steinhoff,
  ``Next-to-next-to-leading order post-Newtonian spin(1)-spin(2) Hamiltonian for self-gravitating binaries,''
  Annalen Phys.\  {\bf 523}, 919 (2011)
  doi:10.1002/andp.201100163
  [arXiv:1107.4294 [gr-qc]].

%\cite{Levi:2011eq}
\bibitem{Levi:2011eq} 
  M.~Levi,
  ``Binary dynamics from spin1-spin2 coupling at fourth post-Newtonian order,''
  Phys.\ Rev.\ D {\bf 85}, 064043 (2012)
  doi:10.1103/PhysRevD.85.064043
  [arXiv:1107.4322 [gr-qc]].

%\cite{Hartung:2013dza}
\bibitem{Hartung:2013dza} 
  J.~Hartung, J.~Steinhoff and G.~Schäfer,
  ``Next-to-next-to-leading order post-Newtonian linear-in-spin binary Hamiltonians,''
  Annalen Phys.\  {\bf 525}, 359 (2013)
  doi:10.1002/andp.201200271
  [arXiv:1302.6723 [gr-qc]].

\bibitem{Levi:2014sba} 
  M.~Levi and J.~Steinhoff,
   ``Equivalence of ADM Hamiltonian and Effective Field Theory approaches at next-to-next-to-leading order spin1-spin2 coupling of binary inspirals,''
  JCAP {\bf 1412}, no. 12, 003 (2014)
  doi:10.1088/1475-7516/2014/12/003
  [arXiv:1408.5762 [gr-qc]].

%\cite{Levi:2015ixa}
\bibitem{Levi:2015ixa}
  M.~Levi and J.~Steinhoff,
   ``Next-to-next-to-leading order gravitational spin-squared potential via the effective field theory for spinning objects in the post-Newtonian scheme,''
  JCAP {\bf 1601}, 008 (2016)
  doi:10.1088/1475-7516/2016/01/008
  [arXiv:1506.05794 [gr-qc]].

\bibitem{Levi:2015msa} 
  M.~Levi and J.~Steinhoff,
   ``Spinning gravitating objects in the effective field theory in the post-Newtonian scheme,''
  JHEP {\bf 1509}, 219 (2015)
  doi:10.1007/JHEP09(2015)219
  [arXiv:1501.04956 [gr-qc]].

\bibitem{Levi:2016ofk} 
  M.~Levi and J.~Steinhoff,
   ``Complete conservative dynamics for inspiralling compact binaries with spins at fourth post-Newtonian order,''
  arXiv:1607.04252 [gr-qc].

%\cite{Levi:2019kgk}
\bibitem{Levi:2019kgk} 
  M.~Levi, S.~Mougiakakos and M.~Vieira,
  ``Gravitational cubic-in-spin interaction at the next-to-leading post-Newtonian order,''
  arXiv:1912.06276 [hep-th].

%\cite{Siemonsen:2019dsu}
\bibitem{Siemonsen:2019dsu} 
  N.~Siemonsen and J.~Vines,
  ``Test black holes, scattering amplitudes and perturbations of Kerr spacetime,''
  arXiv:1909.07361 [gr-qc].

\bibitem{Levi:2014gsa} 
  M.~Levi and J.~Steinhoff,
   ``Leading order finite size effects with spins for inspiralling compact binaries,''
  JHEP {\bf 1506}, 059 (2015)
  doi:10.1007/JHEP06(2015)059
  [arXiv:1410.2601 [gr-qc]].

%\cite{Hergt:2007ha}
\bibitem{Hergt:2007ha} 
  S.~Hergt and G.~Schäfer,
  ``Higher-order-in-spin interaction Hamiltonians for binary black holes from source terms of Kerr geometry in approximate ADM coordinates,''
  Phys.\ Rev.\ D {\bf 77}, 104001 (2008)
  doi:10.1103/PhysRevD.77.104001
  [arXiv:0712.1515 [gr-qc]].

%\cite{Hergt:2008jn}
\bibitem{Hergt:2008jn} 
  S.~Hergt and G.~Schäfer,
  ``Higher-order-in-spin interaction Hamiltonians for binary black holes from Poincare invariance,''
  Phys.\ Rev.\ D {\bf 78}, 124004 (2008)
  doi:10.1103/PhysRevD.78.124004
  [arXiv:0809.2208 [gr-qc]].

%\cite{Vaidya:2014kza}
\bibitem{Vaidya:2014kza} 
  V.~Vaidya,
  ``Gravitational spin Hamiltonians from the S matrix,''
  Phys.\ Rev.\ D {\bf 91}, no. 2, 024017 (2015)
  doi:10.1103/PhysRevD.91.024017
  [arXiv:1410.5348 [hep-th]].

%\cite{Bern:2019nnu}
\bibitem{Bern:2019nnu}
  Z.~Bern, C.~Cheung, R.~Roiban, C.~H.~Shen, M.~P.~Solon and M.~Zeng,
   ``Scattering Amplitudes and the Conservative Hamiltonian for Binary Systems at Third Post-Minkowskian Order,''
  Phys.\ Rev.\ Lett.\  {\bf 122}, no. 20, 201603 (2019)
  doi:10.1103/PhysRevLett.122.201603
  [arXiv:1901.04424 [hep-th]].

%\cite{Bern:2019crd}
\bibitem{Bern:2019crd} 
  Z.~Bern, C.~Cheung, R.~Roiban, C.~H.~Shen, M.~P.~Solon and M.~Zeng,
  ``Black Hole Binary Dynamics from the Double Copy and Effective Theory,''
  JHEP {\bf 1910}, 206 (2019)
  doi:10.1007/JHEP10(2019)206
  [arXiv:1908.01493 [hep-th]].

%\cite{Cheung:2020gyp}
\bibitem{Cheung:2020gyp} 
  C.~Cheung and M.~P.~Solon,
  ``Classical Gravitational Scattering at ${\cal O}(G^3)$ from Feynman Diagrams,''
  arXiv:2003.08351 [hep-th].

%\cite{Blumlein:2020znm}
\bibitem{Blumlein:2020znm} 
  J.~Blümlein, A.~Maier, P.~Marquard and G.~Schäfer,
  ``Testing binary dynamics in gravity at the sixth post-Newtonian level,''
  arXiv:2003.07145 [gr-qc].

%\cite{Vines:2017hyw}
\bibitem{Vines:2017hyw} 
  J.~Vines,
  ``Scattering of two spinning black holes in post-Minkowskian gravity, to all orders in spin, and effective-one-body mappings,''
  Class.\ Quant.\ Grav.\  {\bf 35}, no. 8, 084002 (2018)
  doi:10.1088/1361-6382/aaa3a8
  [arXiv:1709.06016 [gr-qc]].

%\cite{Chung:2020rrz}
\bibitem{Chung:2020rrz} 
  M.~Z.~Chung, Y.~t.~Huang, J.~W.~Kim and S.~Lee,
  ``Complete Hamiltonian for spinning binary systems at first post-Minkowskian order,''
  arXiv:2003.06600 [hep-th].

%\cite{Barack:2018yvs}
\bibitem{Barack:2018yvs} 
  L.~Barack and A.~Pound,
  ``Self-force and radiation reaction in general relativity,''
  Rept.\ Prog.\ Phys.\  {\bf 82}, no. 1, 016904 (2019)
  doi:10.1088/1361-6633/aae552
  [arXiv:1805.10385 [gr-qc]].
	
%\cite{Bini:2018zde}
\bibitem{Bini:2018zde} 
  D.~Bini, T.~Damour, A.~Geralico and C.~Kavanagh,
  ``Detweiler's redshift invariant for spinning particles along circular orbits on a Schwarzschild background,''
  Phys.\ Rev.\ D {\bf 97}, no. 10, 104022 (2018)
  doi:10.1103/PhysRevD.97.104022
  [arXiv:1801.09616 [gr-qc]].
	
%\cite{Bini:2018svh}
\bibitem{Bini:2018svh} 
  D.~Bini and A.~Geralico,
  ``Gravitational self-force corrections to tidal invariants for spinning particles on circular orbits in a Schwarzschild spacetime,''
  Phys.\ Rev.\ D {\bf 98}, no. 8, 084021 (2018)
  doi:10.1103/PhysRevD.98.084021
  [arXiv:1806.03495 [gr-qc]].
	
%\cite{Khalil:2020mmr}
\bibitem{Khalil:2020mmr} 
  M.~Khalil, J.~Steinhoff, J.~Vines and A.~Buonanno,
  ``Fourth post-Newtonian effective-one-body Hamiltonians with generic spins,''
  arXiv:2003.04469 [gr-qc].

%\cite{Nagar:2018plt}
\bibitem{Nagar:2018plt} 
  A.~Nagar, F.~Messina, P.~Rettegno, D.~Bini, T.~Damour, A.~Geralico, S.~Akcay and S.~Bernuzzi,
  ``Nonlinear-in-spin effects in effective-one-body waveform models of spin-aligned, inspiralling, neutron star binaries,''
  Phys.\ Rev.\ D {\bf 99}, no. 4, 044007 (2019)
  doi:10.1103/PhysRevD.99.044007
  [arXiv:1812.07923 [gr-qc]].

%\cite{Teukolsky:1973ha}
\bibitem{Teukolsky:1973ha} 
  S.~A.~Teukolsky,
   ``Perturbations of a rotating black hole. 1. Fundamental equations for gravitational electromagnetic and neutrino field perturbations,''
  Astrophys.\ J.\  {\bf 185}, 635 (1973).
  doi:10.1086/152444
 
\bibitem{Cohen:1974cm}
   J.~M.~Cohen and L.~S.~Kegeles,
    ``Electromagnetic fields in curved spaces - a constructive procedure,''
   Phys.\ Rev.\ D {\bf 10}, 1070 (1974).
   doi:10.1103/PhysRevD.10.1070
 
\bibitem{Chrzanowski:1975wv}
   P.~L.~Chrzanowski,
    ``Vector Potential and Metric Perturbations of a Rotating Black Hole,''
   Phys.\ Rev.\ D {\bf 11}, 2042 (1975).
   doi:10.1103/PhysRevD.11.2042
 
\bibitem{Kegeles:1979an}
   L.~S.~Kegeles and J.~M.~Cohen,
    ``Constructive Procedure For Perturbations Of Space-times,''
   Phys.\ Rev.\ D {\bf 19}, 1641 (1979).
   doi:10.1103/PhysRevD.19.1641

%\cite{Sasaki:2003xr}
\bibitem{Sasaki:2003xr} 
  M.~Sasaki and H.~Tagoshi,
   ``Analytic black hole perturbation approach to gravitational radiation,''
  Living Rev.\ Rel.\  {\bf 6}, 6 (2003)
  doi:10.12942/lrr-2003-6
  [gr-qc/0306120].

%\cite{Keidl:2010pm}
\bibitem{Keidl:2010pm} 
  T.~S.~Keidl, A.~G.~Shah, J.~L.~Friedman, D.~H.~Kim and L.~R.~Price,
  ``Gravitational Self-force in a Radiation Gauge,''
  Phys.\ Rev.\ D {\bf 82}, no. 12, 124012 (2010)
  Erratum: [Phys.\ Rev.\ D {\bf 90}, no. 10, 109902 (2014)]
  doi:10.1103/PhysRevD.82.124012, 10.1103/PhysRevD.90.109902
  [arXiv:1004.2276 [gr-qc]].
	
%\cite{Mano:1996vt}
\bibitem{Mano:1996vt} 
  S.~Mano, H.~Suzuki and E.~Takasugi,
   ``Analytic solutions of the Teukolsky equation and their low frequency expansions,''
  Prog.\ Theor.\ Phys.\  {\bf 95}, 1079 (1996)
  doi:10.1143/PTP.95.1079
  [gr-qc/9603020].

%\cite{Mano:1996mf}
\bibitem{Mano:1996mf} 
  S.~Mano, H.~Suzuki and E.~Takasugi,
   ``Analytic solutions of the Regge-Wheeler equation and the postMinkowskian expansion,''
  Prog.\ Theor.\ Phys.\  {\bf 96}, 549 (1996)
  doi:10.1143/PTP.96.549
  [gr-qc/9605057].

%\cite{Mano:1996gn}
\bibitem{Mano:1996gn} 
  S.~Mano and E.~Takasugi,
   ``Analytic solutions of the Teukolsky equation and their properties,''
  Prog.\ Theor.\ Phys.\  {\bf 97}, 213 (1997)
  doi:10.1143/PTP.97.213
  [gr-qc/9611014].

\bibitem{Regge:1957td} 
  T.~Regge and J.~A.~Wheeler,
   ``Stability of a Schwarzschild singularity,''
  Phys.\ Rev.\  {\bf 108}, 1063 (1957).
  doi:10.1103/PhysRev.108.1063

%\cite{Zerilli:1971wd}
\bibitem{Zerilli:1971wd} 
  F.~J.~Zerilli,
  ``Gravitational field of a particle falling in a Schwarzschild geometry analyzed in tensor harmonics,''
  Phys.\ Rev.\ D {\bf 2}, 2141 (1970).
  doi:10.1103/PhysRevD.2.2141

\bibitem{tulc59}
W.~Tulczyjew, 
``Motion of multipole particles in General Relativity theory,''
 Acta \ Phys.\ Polon. {\bf 18}, 393 (1959).

%\cite{Dixon:1970zz}
\bibitem{Dixon:1970zz} 
  W.~G.~Dixon,
  ``Dynamics of extended bodies in general relativity. II. Moments of the charge-current vector,''
  Proc.\ Roy.\ Soc.\ Lond.\ A {\bf 319}, 509 (1970).
  doi:10.1098/rspa.1970.0191

\bibitem{ehlers77}
J. Ehlers and E. Rudolph,
``Dynamics of extended bodies in general relativity center-of-mass description and quasirigidity,''
Gen.\ Rel.\  Grav. {\bf 8},197-217 (1977).
doi: 10.1007/BF00763547 

\bibitem{Thorne:1980ru} 
  K.~S.~Thorne,
   ``Multipole Expansions of Gravitational Radiation,''
  Rev.\ Mod.\ Phys.\  {\bf 52}, 299 (1980).
  doi:10.1103/RevModPhys.52.299

%\cite{Shah:2012gu}
\bibitem{Shah:2012gu} 
  A.~G.~Shah, J.~L.~Friedman and T.~S.~Keidl,
  ``EMRI corrections to the angular velocity and redshift factor of a mass in circular orbit about a Kerr black hole,''
  Phys.\ Rev.\ D {\bf 86}, 084059 (2012)
  doi:10.1103/PhysRevD.86.084059
  [arXiv:1207.5595 [gr-qc]].

%\cite{Blanchet:2012at}
\bibitem{Blanchet:2012at} 
  L.~Blanchet, A.~Buonanno and A.~Le Tiec,
  ``First law of mechanics for black hole binaries with spins,''
  Phys.\ Rev.\ D {\bf 87}, no. 2, 024030 (2013)
  doi:10.1103/PhysRevD.87.024030
  [arXiv:1211.1060 [gr-qc]].

%\cite{Porto:2008jj}
\bibitem{Porto:2008jj} 
  R.~A.~Porto and I.~Z.~Rothstein,
  ``Next to Leading Order Spin(1)Spin(1) Effects in the Motion of Inspiralling Compact Binaries,''
  Phys.\ Rev.\ D {\bf 78}, 044013 (2008)
  Erratum: [Phys.\ Rev.\ D {\bf 81}, 029905 (2010)]
  doi:10.1103/PhysRevD.81.029905, 10.1103/PhysRevD.78.044013
  [arXiv:0804.0260 [gr-qc]].

%\cite{Barker:1975ae}
\bibitem{Barker:1975ae} 
  B.~M.~Barker and R.~F.~O'Connell,
  ``Gravitational Two-Body Problem with Arbitrary Masses, Spins, and Quadrupole Moments,''
  Phys.\ Rev.\ D {\bf 12}, 329 (1975).
  doi:10.1103/PhysRevD.12.329

%\cite{DEath:1975wqz}
\bibitem{DEath:1975wqz} 
  P.~D.~D'Eath,
  ``Interaction of two black holes in the slow-motion limit,''
  Phys.\ Rev.\ D {\bf 12}, 2183 (1975).
  doi:10.1103/PhysRevD.12.2183

\bibitem{Barker:OConnell:1979}
  B.~M.~Barker and R.~F.~O'Connell,
  ``The gravitational interaction: Spin, rotation, and quantum effects---a review,''
  Gen.\ Relativ.\ Gravit.\ {\bf 11} 149 (1979).
  doi:10.1007/BF00756587

%\cite{Thorne:1984mz}
\bibitem{Thorne:1984mz} 
  K.~S.~Thorne and J.~B.~Hartle,
  ``Laws of motion and precession for black holes and other bodies,''
  Phys.\ Rev.\ D {\bf 31}, 1815 (1984).
  doi:10.1103/PhysRevD.31.1815

%\cite{Steinhoff:2007mb}
\bibitem{Steinhoff:2007mb} 
  J.~Steinhoff, S.~Hergt and G.~Schäfer,
  ``On the next-to-leading order gravitational spin(1)-spin(2) dynamics,''
  Phys.\ Rev.\ D {\bf 77}, 081501 (2008)
  doi:10.1103/PhysRevD.77.081501
  [arXiv:0712.1716 [gr-qc]].

%\cite{Porto:2006bt}
\bibitem{Porto:2006bt} 
  R.~A.~Porto and I.~Z.~Rothstein,
  ``The Hyperfine Einstein-Infeld-Hoffmann potential,''
  Phys.\ Rev.\ Lett.\  {\bf 97}, 021101 (2006)
  doi:10.1103/PhysRevLett.97.021101
  [gr-qc/0604099].

%\cite{Porto:2008tb}
\bibitem{Porto:2008tb} 
  R.~A.~Porto and I.~Z.~Rothstein,
  ``Spin(1)Spin(2) Effects in the Motion of Inspiralling Compact Binaries at Third Order in the Post-Newtonian Expansion,''
  Phys.\ Rev.\ D {\bf 78}, 044012 (2008)
  Erratum: [Phys.\ Rev.\ D {\bf 81}, 029904 (2010)]
  doi:10.1103/PhysRevD.78.044012, 10.1103/PhysRevD.81.029904
  [arXiv:0802.0720 [gr-qc]].

%\cite{Levi:2008nh}
\bibitem{Levi:2008nh} 
  M.~Levi,
  ``Next to Leading Order gravitational Spin1-Spin2 coupling with Kaluza-Klein reduction,''
  Phys.\ Rev.\ D {\bf 82}, 064029 (2010)
  doi:10.1103/PhysRevD.82.064029
  [arXiv:0802.1508 [gr-qc]].

%\cite{Steinhoff:2008ji}
\bibitem{Steinhoff:2008ji} 
  J.~Steinhoff, S.~Hergt and G.~Schäfer,
  ``Spin-squared Hamiltonian of next-to-leading order gravitational interaction,''
  Phys.\ Rev.\ D {\bf 78}, 101503 (2008)
  doi:10.1103/PhysRevD.78.101503
  [arXiv:0809.2200 [gr-qc]].

%\cite{Hergt:2010pa}
\bibitem{Hergt:2010pa} 
  S.~Hergt, J.~Steinhoff and G.~Schäfer,
  ``Reduced Hamiltonian for next-to-leading order Spin-Squared Dynamics of General Compact Binaries,''
  Class.\ Quant.\ Grav.\  {\bf 27}, 135007 (2010)
  doi:10.1088/0264-9381/27/13/135007
  [arXiv:1002.2093 [gr-qc]].

%\cite{Hergt:2011ik}
\bibitem{Hergt:2011ik} 
  S.~Hergt, J.~Steinhoff and G.~Schäfer,
  ``Elimination of the spin supplementary condition in the effective field theory approach to the post-Newtonian approximation,''
  Annals Phys.\  {\bf 327}, 1494 (2012)
  doi:10.1016/j.aop.2012.02.006
  [arXiv:1110.2094 [gr-qc]].

%\cite{Bohe:2015ana}
\bibitem{Bohe:2015ana} 
  A.~Bohé, G.~Faye, S.~Marsat and E.~K.~Porter,
  ``Quadratic-in-spin effects in the orbital dynamics and gravitational-wave energy flux of compact binaries at the 3PN order,''
  Class.\ Quant.\ Grav.\  {\bf 32}, no. 19, 195010 (2015)
  doi:10.1088/0264-9381/32/19/195010
  [arXiv:1501.01529 [gr-qc]].

%\cite{Levi:2015uxa}
\bibitem{Levi:2015uxa} 
  M.~Levi and J.~Steinhoff,
  ``Next-to-next-to-leading order gravitational spin-orbit coupling via the effective field theory for spinning objects in the post-Newtonian scheme,''
  JCAP {\bf 1601}, 011 (2016)
  doi:10.1088/1475-7516/2016/01/011
  [arXiv:1506.05056 [gr-qc]].

\bibitem{Hergt:2012zx} 
  S.~Hergt, J.~Steinhoff and G.~Schäfer,
   ``On the comparison of results regarding the post-Newtonian approximate treatment of the dynamics of extended spinning compact binaries,''
  J.\ Phys.\ Conf.\ Ser.\  {\bf 484}, 012018 (2014)
  doi:10.1088/1742-6596/484/1/012018
  [arXiv:1205.4530 [gr-qc]].


%\cite{Bini:2019lcd}
\bibitem{Bini:2019lcd} 
  D.~Bini and A.~Geralico,
  ``New gravitational self-force analytical results for eccentric equatorial orbits around a Kerr black hole: redshift invariant,''
  Phys.\ Rev.\ D {\bf 100}, no. 10, 104002 (2019)
  doi:10.1103/PhysRevD.100.104002
  [arXiv:1907.11080 [gr-qc]].
	

%\cite{Balmelli:2015zsa}
\bibitem{Balmelli:2015zsa} 
  S.~Balmelli and T.~Damour,
  ``New effective-one-body Hamiltonian with next-to-leading order spin-spin coupling,''
  Phys.\ Rev.\ D {\bf 92}, no. 12, 124022 (2015)
  doi:10.1103/PhysRevD.92.124022
  [arXiv:1509.08135 [gr-qc]].

%\cite{Damour:2001tu}
\bibitem{Damour:2001tu} 
  T.~Damour,
  ``Coalescence of two spinning black holes: an effective one-body approach,''
  Phys.\ Rev.\ D {\bf 64}, 124013 (2001)
  doi:10.1103/PhysRevD.64.124013
  [gr-qc/0103018].

%\cite{Barausse:2009xi}
\bibitem{Barausse:2009xi} 
  E.~Barausse and A.~Buonanno,
  ``An Improved effective-one-body Hamiltonian for spinning black-hole binaries,''
  Phys.\ Rev.\ D {\bf 81}, 084024 (2010)
  doi:10.1103/PhysRevD.81.084024
  [arXiv:0912.3517 [gr-qc]].

%\cite{Bini:2019nra}
\bibitem{Bini:2019nra} 
  D.~Bini, T.~Damour and A.~Geralico,
  ``Novel approach to binary dynamics: application to the fifth post-Newtonian level,''
  Phys.\ Rev.\ Lett.\  {\bf 123}, no. 23, 231104 (2019)
  doi:10.1103/PhysRevLett.123.231104
  [arXiv:1909.02375 [gr-qc]].

%\cite{Damour:2019lcq}
\bibitem{Damour:2019lcq} 
  T.~Damour,
  ``Classical and Quantum Scattering in Post-Minkowskian Gravity,''
  arXiv:1912.02139 [gr-qc].

%\cite{Levi:2020uwu}
\bibitem{Levi:2020uwu} 
  M.~Levi, A.~J.~McLeod and M.~von Hippel,
  ``NNNLO gravitational quadratic-in-spin interactions at the quartic order in G,''
  arXiv:2003.07890 [hep-th].

%\cite{Foffa:2019hrb}
\bibitem{Foffa:2019hrb} 
  S.~Foffa, P.~Mastrolia, R.~Sturani, C.~Sturm and W.~J.~Torres Bobadilla,
  ``Static two-body potential at fifth post-Newtonian order,''
  Phys.\ Rev.\ Lett.\  {\bf 122}, no. 24, 241605 (2019)
  doi:10.1103/PhysRevLett.122.241605
  [arXiv:1902.10571 [gr-qc]].

%\cite{Blumlein:2019zku}
\bibitem{Blumlein:2019zku} 
  J.~Blümlein, A.~Maier and P.~Marquard,
  ``Five-Loop Static Contribution to the Gravitational Interaction Potential of Two Point Masses,''
  Phys.\ Lett.\ B {\bf 800}, 135100 (2020)
  doi:10.1016/j.physletb.2019.135100
  [arXiv:1902.11180 [gr-qc]].

	
\end{thebibliography}
\end{document}